\documentclass{emulateapj}

\def\bmath#1{\mbox{\boldmath $#1$}}

\slugcomment{}
\shorttitle{Relativistic Radiation HD}
\shortauthors{Takahashi et al.}
\begin{document}

\title{Explicit-Implicit Scheme for Relativistic Radiation
Hydrodynamics}

\author{Hiroyuki R. Takahashi\altaffilmark{1}, 
Ken Ohsuga\altaffilmark{2}, Yuichiro Sekiguchi \altaffilmark{3},
Tsuyoshi Inoue\altaffilmark{4}, and Kengo Tomida\altaffilmark{2, 5}}
\email{takahashi@cfca.jp}
\altaffiltext{1}{Center for Computational Astrophysics, National
  Astronomical Observatory of Japan, Mitaka, Tokyo 181-8588, Japan}
\altaffiltext{2}{Division of Theoretical Astronomy, National
  Astronomical Observatory of Japan, Mitaka, Tokyo 181-8588, Japan}
\altaffiltext{3}{Yukawa Institute for Theoretical Physics, Kyoto University, Kyoto 606-8502, Japan}
\altaffiltext{4}{Department of Physics and Mathematics, Aoyama Gakuin University, Fuchinobe, Sagamihara 229-8558, Japan}
\altaffiltext{5}{Department of Astronomical Science, The Graduate University for Advanced Studies (SOKENDAI), Osawa, Mitaka, Tokyo 181-8588, Japan}




\begin{abstract}
We propose an explicit-implicit scheme for numerically solving Special
Relativistic Radiation Hydrodynamic (RRHD) equations, which
ensures a conservation of total energy and momentum (matter and
radiation). In our scheme, 
0th and 1st moment equations of the radiation transfer equation
are numerically solved without employing a flux-limited 
diffusion (FLD) approximation. 
For an hyperbolic term, 
of which the time scale is the light crossing time 
when the flow velocity is comparable to the speed of light,
is explicitly solved using an approximate Riemann solver. 
Source terms describing an exchange of energy and momentum 
between the matter and the radiation via the gas-radiation interaction
are implicitly integrated using an iteration method. 
The implicit scheme allows us to relax the Courant-Friedrichs-Lewy
condition in optically thick media, where heating/cooling and scattering
 timescales could be much shorter than the dynamical timescale.
We show that our numerical code can pass test problems of 
one- and two-dimensional radiation energy transport,
and one-dimensional radiation hydrodynamics.
Our newly developed scheme could be useful for a
number of relativistic astrophysical problems.
We also discuss how to extend our explicit-implicit scheme to the
 relativistic radiation magnetohydrodynamics.
\end{abstract}
\section{Introduction}\label{intro}
Relativistic flows appear in many high-energy astrophysical phenomena,
such as jets from microquasars and active galactic nuclei (AGNs),
pulsar winds, magnetar flares, core collapse supernovae, and gamma-ray
bursts (GRBs). In many of these systems, the magnetic field has a
crucial role in dynamics. For example, magnetic fields
connecting an accretion disk with a central
star, or, different points of accretion disks are twisted and amplified
due to the differential rotation,
launching the jets \citep{1976Natur.262..649L, 1982MNRAS.199..883B,
1985PASJ...37..515U}. The production of astronomical jets has been
observed in magnetohydrodynamic (MHD) simulations \citep{1996ApJ...468L..37H,
2001Sci...291...84M, 2004ApJ...600..338K}.
The jets are also powered by a subtraction of the rotational energy of a
central black hole via the magnetic field \citep{1977MNRAS.179..433B,
2002Sci...295.1688K,
2009MNRAS.394L.126M}. Also the magnetic fields play an important role in
accretion disks to transport the angular momentum outward, leading to
the mass accretion \citep{1991ApJ...376..214B, 1995ApJ...440..742H}

Not merely the observational point of view, but the radiation field
is also an important ingredient in the dynamics of relativistic phenomena. 
The radiation pressure force
would play a key role in jet acceleration 
\citep{1977A&A....59..111B, 1996MNRAS.280..781S, 2005ApJ...628..368O, 2009MNRAS.398.1668O}. 
Recently,
\cite{2010PASJ...62L..43T} showed a formation of 
radiatively accelerated and magnetically collimated jets
using non-relativistic radiation MHD (RMHD) simulations 
\citep[see also][]{2009PASJ...61L...7O, 2011ApJ...736....2O}.
The radiative effect also plays important roles 
in non-relativistic phenomena.
For example, the dominance of the radiation pressure in optically thick,
geometrically thin (standard) accretion disks
could be unstable to thermal and viscous instabilities
\citep{1974ApJ...187L...1L, 1975PThPh..54..706S, 1976MNRAS.175..613S,
2011ApJ...727..106T}.
Here we note that the global radiation hydrodynamic (RHD) simulations
showed the growth of instabilities \citep{2006ApJ...640..923O},
but \cite{2009ApJ...704..781H} concluded 
that the radiation pressure dominated region
is thermally stable using RMHD simulations
of a local patch of the disk.
Thus the high energy phenomena would be understood only after we take
into account the magnetic field and the radiation consistently.

Many methods have been proposed to include radiative effects in
non-relativistic plasmas. Solving a full radiation transfer equation is
the most challenging task for directly connecting results between 
numerical simulations and observations \citep{1992ApJS...80..819S}. 
But it is a hard task to couple with the hydrodynamical code due to its
complexity and high computational costs.
Another class of simple method is based on the moment method. 
In the Flux-Limited Diffusion approximation (FLD),
a 0th moment equation of the radiation transfer 
equation is solved to evaluate the radiation energy density, $E'_r$.
The radiative flux, $\bmath F'_r$, is given 
based on the gradient of the radiation energy density
without solving the 1st moment equation.
This is quite a useful technique and gives appropriate
radiation fields within the optically thick regime,
but we should keep in mind
that it does not always give precise radiation fields 
in the regime where the optical depth
is around unity or less \citep[see][]{2011ApJ...736....2O}.
Thus it is better to solve the 0th and 1st moment equations.

When we solve two moment equations,
the closure relation, $P'^{ij}_r=D^{ij} E_r'$,
is used to evaluate the radiation stress, $P_r'^{ij}$,
where $D^{ij}$ is the Eddington tensor.
In the Eddington approximation,
the tensor is simply given by 
$D^{ij}= \delta^{ij}/3$, where $\delta^{ij}$ is the Kronecker delta. 
This implies that the radiation field is isotropic 
in the comoving frame, which is a reasonable assumption in the
optically thick media.
Another class of the Eddington tensor is obtained by approximately
taking into account the anisotropy of radiation fields 
\citep[so called M-1 closure, ][]{1976UCRL...78378L.CA,
1978JQSRT..20..541M, 1984JQSRT..31..149L}.
\cite{2007A&A...464..429G} implemented the M-1 closure in their
non-relativistic hydrodynamic code 
and demonstrated that the anisotropic radiation fields
around an obstacle are accurately solved.

In the framework of relativistic radiation hydrodynamics (RRMHD), 
\cite{2008PhRvD..78b4023F} first proposed a numerical scheme for solving
general relativistic radiation
magnetohydrodynamic equations assuming an optically thick
medium (i.e., $D^{ij}=\delta^{ij}/3$). \cite{2011MNRAS.tmp.1386Z}
implemented the general
relativistic radiation hydrodynamics code and adopted it for the
Bondi-Hoyle accretion on to the black hole. 
Recently, \cite{2011PThPh.125.1255S} proposed another truncated moment
formalism of radiation fields in optically thick and thin limits. 

In the RR(M)HD, there are three independent timescales.
The first one is the dynamical time scale $t_{dyn}\equiv L/v$, 
where $L$ and $v$ are the typical size of the system and the fluid
velocity. 
The second one is the timescale at which the characteristic wave passes
the system,
$t_w\equiv L/\lambda$, where $\lambda$ is the characteristic wave
speed. 
These two timescales are comparable and close to the light crossing
time, $L/c$ (where $c$ is the speed of light) in
relativistic phenomena (i.e., $v\sim \lambda \sim c$).
The third one is the heating/cooling and scattering timescales, 
$t_{\rm ab}$ and $t_{\rm sc}$. These timescales 
become much shorter than the dynamical one in the optically thick medium,
since the gas and the radiation are frequently interact.
This indicates that equations for the radiation fields become stiff. 
In the explicit scheme, a numerical time step $\Delta t$ should be
restricted to being shorter than these typical timescales 
($\Delta t < \min[t_{dyn},t_{w},t_{ab},t_{sc}]$) 
to ensure the numerical stability. 
Therefore, high computational costs prevent us from studying long
term evolutions when the gas is optically thick,
if we are interested in the phenomena of the dynamical timescale.

In this paper, we propose an explicit-implicit numerical scheme overcome
this problem. As the first step, we consider the relativistic radiation
hydrodynamics by neglecting the magnetic field (but, see section
\ref{discussion} for the discussion how to extend our numerical scheme to
RRMHD).
The proposed scheme ensures a conservation of total energy and
momentum. 
Governed equations are integrated in time using both explicit and
implicit schemes.
The former solves an hyperbolic term 
(timescale $\sim t_{w}\sim t_{dyn}$). The latter treats the
exchange of energy and momentum between the radiation and the matter
via the gas-radiation interaction,
whose time scales are $\sim t_{ab}$ and $t_{sc}$.
This method allows us to take a larger time step 
$\Delta t > t_{ab}, t_{sc}$ than that of the explicit scheme.

This paper is organized as follows: In \S~\ref{Method}, we introduce
argument equations for RRHD and the numerical scheme is shown in
\S~\ref{Numerical}. Numerical results of one- and two-dimensional tests
are shown in \S~\ref{test}. Discussion and summary are appeared in
\S~\ref{discussion} and \S~\ref{summary}.

\section{Basic Equations}\label{Method}
In the following, we take the light speed as unity. 
The special relativistic radiation magnetohydrodynamic equations of
ideal gas consist of conservation of mass,
\begin{equation}
 (\rho u^\nu)_{,\nu} = 0,\label{geq:mcons}
\end{equation}
conservation of energy-momentum,
\begin{equation}
\left(T^{\mu\nu}_\mathrm{HD} 
+T^{\mu\nu}_\mathrm{rad}\right)_{,\nu} =0,\label{geq:Tcons}
\end{equation}
and equations of radiation energy-momentum
\begin{equation}
 T^{\mu\nu}_\mathrm{rad,\nu} = -G^\mathrm{\mu},\label{geq:Tradcons}
\end{equation}
where $\rho$ is the proper mass density, $u^\mu = \gamma(1 , v^i)$
is the fluid four velocity, and $T^{\mu\nu}_\mathrm{HD}$ and
$T^{\mu\nu}_\mathrm{rad}$ are energy momentum tensors of fluid
and radiation. Here $\gamma = \sqrt{1+u_i u^i}$ is the bulk Lorentz
factor and $v^i$ is the fluid three velocity. Greek indices range over $0, 1, 2, 3$ and Latin ranges over
$1, 2, 3$, where $0$ indicates the time component and $1, 2, 3$ do space
components. 

The energy momentum tensor of fluids is written as 
\begin{equation}
  T^{\mu\nu}_\mathrm{HD} = \rho \xi u^\mu u^\nu + p_g  \eta^{\mu \nu},\label{geq:THD}
\end{equation}
where $p_g$ is the gas pressure and
$\mathrm{diag}~\eta^{\mu\nu}=(-1,1,1,1)$ is the Minkowski metric. The
specific enthalpy of relativistic ideal gas $\xi$ is given by
\begin{equation}
 \xi = 1 + \frac{\Gamma}{\Gamma-1}\frac{p_g}{\rho},\label{geq:xi}
\end{equation}
where $\Gamma$ is the specific heat ratio. 

The energy momentum tensor of radiation is written as
\begin{equation}
 T^{\mu\nu}_\mathrm{rad}
  =\left(\begin{array}{cc}
    E_r, & F_r^j \\
	   F_r^{i}, & P^{ij}_r\label{geq:Trad}
	 \end{array}\right)
\end{equation}
where $E_r$, $F_r^i$, and $P^{ij}_r$ are the radiation energy
density, flux and stress measured in the laboratory frame.

The radiation exchanges its energy and momentum with fluids through
absorption/emission and scattering processes. The radiation four force
$G^\mu$ is explicitly given by
\begin{eqnarray}
 G^0 &=& -\rho \kappa 
  \left(4\pi \mathrm{B} \gamma - \gamma E_r + u_i F_r^i\right)
  \nonumber \\
 &-& \rho \sigma_s\left[\gamma u^2 E_r + \gamma u_i u_j
	      P^{ij}_r-\left(\gamma^2+u^2\right)
	      u_i F_r^i\right],\label{geq:G0}
\end{eqnarray}
and 
\begin{eqnarray}
 G^{i} &=&- 4\pi \rho \kappa \mathrm{B} u^i 
 + \rho (\kappa + \sigma_s)(\gamma F_r^i-u_jP^{ij}_r) \nonumber \\
 &-&\rho \sigma_s u^i
\left(\gamma^2 E_r - 2\gamma u_j F_r^j + u_j u_k P_r^{jk}\right),\label{geq:Gi}
\end{eqnarray}
where $u = \sqrt{u_i u^i}$. 
$\kappa$ and $\sigma_s$ are absorption and scattering
coefficients measured in the comoving frame. Thus, equation
(\ref{geq:Tradcons}) is a mixed-frame radiation energy-momentum
equation that the radiation field is defined in the observer frame,
while the absorption and scattering coefficients are defined in the
comoving frame. 

In equations (\ref{geq:G0}) and (\ref{geq:Gi}), we assume the
Kirchhoff-Planck relation so that the emissivity
$\eta$ is replaced by the blackbody intensity $\mathrm{B}$ as
$\eta=\kappa \mathrm{B}$ in the comoving frame.

The blackbody intensity $\mathrm{B}$ is a function of gas temperature
$T$ as $\mathrm{B}=a_R T^4/(4\pi)$, where $a_R$ is related to the
Stefan-Boltzmann constant
$\sigma_\mathrm{SB}=a_R/4$. The plasma temperature is determined by
\begin{equation}
p_g =  \frac{\rho k_B T}{\mu m_p},\label{geq:eos}
\end{equation}
where $k_B$ and $m_p$ are the Boltzmann constant and the proton mass,
and $\mu$ is the mean molecular weight.

Since we consider the moment equation of radiation fields, we need
another relation between $E_r$, $F_r^i$ and $P_r^{ij}$ to close
the system, i.e., the closure relation. Here we leave it as a general
form described by
\begin{equation}
 P'^{ij}_r = P'^{ij}(E_r', F_r'^k),\label{geq:closure}
\end{equation}
where dash denotes the quantity defined in the comoving frame. 
The explicit form of closure relation is introduced in \S~\ref{closure}
after the formulation.


In the following, we deal with the one-dimensional conservation law in
the $x$-direction:
\begin{equation}
 \frac{\partial \mathcal{U}}{\partial t} + \frac{\partial
  \mathcal{F}}{\partial x} = \mathcal{S},\label{geq:1dform}
\end{equation}
which follows equations (\ref{geq:mcons}), (\ref{geq:Tcons}), and 
(\ref{geq:Tradcons}).
Primitive variables $\mathcal{P}$ are defined as
\begin{equation}
 \mathcal{P} = 
  \left(\begin{array}{c}
   \rho \\
   u^k \\
   p_g \\
   E_r \\
   F_r^k\\
\end{array}\right).
\end{equation}
Conserved variables
$\mathcal{U}$, fluxes $\mathcal{F}$, and source terms $\mathcal{S}$ have
forms of
\begin{equation}
 \mathcal{U} 
  = \left(\begin{array}{c}
     D \\
	   m_t^k\\
     E_t\\
     E_r\\
     F_r^k\\
  \end{array}\right),
\end{equation}
\begin{equation}
  \mathcal{F}
  = \left(\begin{array}{c}  
     D v^x\\
     \rho \xi u^x u^k + p_g\delta^{xk} + P_r^{xk}\\
     m_t^x \\
     F_r^x \\
     P_r^{xk}
   \end{array}\right),\label{geq:Flux}
\end{equation}
and 
\begin{equation}
 \mathcal{S} 
  \equiv \left(\begin{array}{c}
     0 \\
     0\\
     0\\
     S_E\\
     S_F^k\\
  \end{array}\right) 
=
 \left(\begin{array}{c}
     0 \\
     0\\
     0\\
     -G^0\\
     -G^k\\
  \end{array}\right),\label{def:source}
\end{equation}
where $D=\rho \gamma$ is the mass density measured in the laboratory
frame.
The total energy density $E_t$ and momentum density $m_t$ are given by
\begin{equation}
 E_t = E_\mathrm{HD} + E_r = \rho \xi \gamma^2 - p_g + E_r,\label{geq:Et}
\end{equation}
and 
\begin{equation}
 m^k_t = m^k_\mathrm{HD} + F_r^k = \rho \xi \gamma u^k  + F_r^k,\label{geq:mt}
\end{equation}
where $E_\mathrm{HD}$ and $m^k_\mathrm{HD}$ denote the energy and momentum
density of the fluids, respectively. 

It should be noted that $E_r$ and $F_r^k$ are not only the primitive
variables, but also the conserved variables. Thus, it is straightforward
for the radiation fields to convert the primitive variables from the
conserved variables, as we will see in \S~\ref{sc:source}.

\section{Numerical Scheme for RRHD}\label{Numerical}
In this section, we propose a new numerical scheme for solving RRHD
equations. 
A conservative discretization of 1-dimensional equation
(\ref{geq:1dform}) over a time step $\Delta t$ from $t=n \Delta t$ to $t
= (n+1)\Delta t$ with grid spacing $\Delta x$ is written as
\begin{equation}
 \mathcal{U}^{n+1}_i= \mathcal{U}^n_i - \frac{\Delta t}{\Delta x}
  \left(f_{i+\frac{1}{2}} - f_{i-\frac{1}{2}}\right) +
  \mathcal{S}_i\Delta t,
  \label{eq:1ddif}
\end{equation}
where $\mathcal{U}^n_i$ is the conservative variable at $x=x_i$ and
$t=n\Delta t$, and $f_{i\pm \frac{1}{2}}$ is the upwind numerical flux at the cell
surfaces $x=x_{i\pm \frac{1}{2}}$. In the numerical procedure, we divide
equation (\ref{eq:1ddif}) into two parts, the {\it hyperbolic term } 
\begin{equation}
 \mathcal{U}^{*}_i = \mathcal{U}^{n}_i 
  - \frac{\Delta t}{\Delta x}
  \left(f_{i+\frac{1}{2}} - f_{i-\frac{1}{2}}\right),\label{geq:fluxterm}
\end{equation}
and the {\it source term} 
\begin{equation}
 \mathcal{U}^{n+1}_i = \mathcal{U}^{*}_i + \mathcal{S}_i \Delta t, \label{geq:sourceterm}
\end{equation}
where $\mathcal{U}^{*}$ is the conservative variable at the auxiliary step. 
In the following subsection, we describe how to solve these two
equations. 

\subsection{hyperbolic term}
For the hyperbolic term, equation (\ref{geq:fluxterm}) is solved as 
an initial value problem with the initial condition
\begin{equation}
 \mathcal{U}(t^n, x) = \left\{
\begin{array}{lll}
 \mathcal{U}_L & \mathrm{for} & x<x_{i+\frac{1}{2}}\\
 \mathcal{U}_R & \mathrm{for} & x>x_{i+\frac{1}{2}}
\end{array}
\right.
\end{equation}
where the subscript $L$ and $R$ denote left and right constant
states on the cell interface. 
When we take $\mathcal{U}_L = \mathcal{U}_i$ and $\mathcal{U}_R =
\mathcal{U}_{i+1}$, the numerical solution is reliable with a first order
accuracy in space. 

In order to improve the spatial accuracy, primitive variables
at the zone surface $\mathcal{P}_{i\pm 1/2}$ are calculated by
interpolating them from a cell center to a cell surface (the so-called
reconstruction step). The primitive
variables at the left and right states are written as
\begin{eqnarray}
 \mathcal{P}_{i\pm \frac{1}{2},S} = \mathcal{P}_i 
  \pm \frac{\delta_x \mathcal{P}}{2},
\end{eqnarray}
where we take $S=L (R)$ with the plus (minus) sign. 
The slope $\delta_x P$ should be determined so as to preserve the
monotonicity. In our numerical code, we
adopt the harmonic mean proposed by 
\cite{1977JCoPh..23..263V}, which has a 2nd order accuracy in space, as
\begin{equation}
 \delta_x \mathcal{P}= 
\frac{2 \mathrm{max(0,\Delta \mathcal{P}_+ \Delta \mathcal{P}_-)}}
{\Delta \mathcal{P}_+ + \Delta \mathcal{P}_-},
\end{equation}
where
\begin{equation}
 \Delta \mathcal{P}_\pm = \pm (\mathcal{P}_{i\pm1} - \mathcal{P}_i).
\end{equation}
Here we adopted a 2nd order accurate scheme, but a higher order
scheme such as Piecewise Constant Method \citep{1999MNRAS.308.1069K},
Piecewise Parabolic Method \citep{1984JCoPh..54..174C,
1996JCoPh.123....1M} or the other scheme can be applicable to both hydropart
and radiation part.

By computing the primitive variables at the cell surface, the flux
$\mathcal{F}_{i \pm 1/2,S}$ is calculated directly from
$\mathcal{P}_{i\pm\frac{1}{2},S}$ 
(but see \S~\ref{closure} for the procedure of calculating $P_r^{xk}$
appeared in equation \ref{geq:Flux} from $P'^{ij}_r$).
Then, numerical fluxes are calculated using an approximate Riemann
solver. 
We adopted the HLL method \citep{1983siamRev...25..35..61}, which
can capture the propagation of fastest waves. The numerical flux is then
computed as
\begin{eqnarray}
f_{i+\frac{1}{2}}=
 \frac{\lambda_+\mathcal{F}_{i+\frac{1}{2},L} 
- \lambda_-\mathcal{F}_{i+\frac{1}{2},R} 
+\lambda_+\lambda_-
  (\mathcal{U}_{i+\frac{1}{2},R} - \mathcal{U}_{i+\frac{1}{2},L})}
{\lambda_+ - \lambda_-}.\nonumber \\
\label{eq:hllf}
\end{eqnarray}
Here $\lambda_-$ and $\lambda_+$ are
\begin{equation}
 \lambda_-= \mathrm{min}(0, \lambda_{L-}, \lambda_{R-}),
\end{equation}
and
\begin{equation}
  \lambda_+= \mathrm{max}(0, \lambda_{L+}, \lambda_{R+}),
\end{equation}
where $\lambda_{S+}$ and $\lambda_{S-}$ are the right and left going
wave speed of the fastest mode. For example, they correspond to the
sound speeds for the relativistic hydrodynamics. When the radiation
field is included, the light mode determines the fastest wave speed,
which depends on the closure relation. The fastest wave speed is
discussed after specifying the closure relation in \S~\ref{closure}.
We note that although we adopted the HLL scheme for simplicity, the
higher order approximate Riemann solvers such as the HLLC
\citep{2005MNRAS.364..126M, 2006MNRAS.368.1040M, 2007JCoPh.223..643H}
and HLLD \citep{2009MNRAS.393.1141M} can be
implemented in the relativistic radiation (Magneto)hydrodynamics
\citep{2009camcs4.135}.

Using the numerical flux $f_{\pm \frac{1}{2}}$, the conserved variables
$U^*$ is obtained from equation (\ref{geq:fluxterm}).
It should be noted that the total energy density $E_t^{n+1}$ and
momentum density $m_t^{n+1,k}$ at $t=(n+1)\Delta t$ are already
solved although we only integrate the hyperbolic term. 
Thus, when the radiation energy density and flux are obtained by solving
equation (\ref{geq:sourceterm}), the fluid energy density and momentum
density are immediately computed (see, equations \ref{eq:Emhd}-\ref{eq:Fmhd}). 



\subsection{Source term}\label{sc:source}
Next, we show how to solve equation (\ref{geq:sourceterm}). 
The source term appears in equation (\ref{geq:Tradcons}),
which treats the interaction between the radiation and the matter.
As discussed in \S~\ref{intro}, the heating/cooling and scattering timescales
can be shorter than the dynamical timescale in optically thick media. 
It prevents us from studying the long time evolution ($\sim t_{dyn}$)
when the equation is explicitly integrated. 
To overcome this difficulty, we want to construct an implicit scheme as
\begin{eqnarray} 
 E_r^{n+1} = E_r^{*} + \Delta t \mathcal{S}_E(E_r^{n+1}, 
  {F}_r^{n+1,j}, P_r^{n+1,jk},  \mathcal{P}_h^{n+1}),\nonumber \\
{\label{eq:gimp1-1}}\\
 F_r^{n+1,i} = F_r^{*,i} + \Delta t \mathcal{S}_F^i(E_r^{n+1}, 
  {F}_r^{n+1,j}, P_r^{n+1,jk},
  \mathcal{P}_h^{n+1}),{\label{eq:gimp1-2}}\nonumber \\
\end{eqnarray}
where $\mathcal{P}_h$ is primitive variables of fluids
(i.e., $\rho, u^i, p_g$). 
We confront two problems to solve equations ({\ref{eq:gimp1-1}}) -
({\ref{eq:gimp1-2}}).
The first problem comes from the appearance of $\mathcal{P}_h^{n+1}$.
Since $\mathcal{P}_h^{n+1}$ should be obtained after computing
$\mathcal{U}^{n+1}$,
equations (\ref{eq:gimp1-1}) - (\ref{eq:gimp1-2}) become non-linear
equations for $\mathcal{U}^{n+1}$.
Another difficulty comes from the closure relation $P^{ij}_r = D^{ij}
E_r$. The Eddington tensor
$D^{ij}=D^{ij}(E_r, F_r^i)$ is generally a non-linear function of
$E_r$ and $F_r^i$, so we cannot adopt the simple implicit method by
replacing $E_r\rightarrow E_r^{n+1}$ and $F_r^i\rightarrow
F_r^{n+1,i}$ in the Eddington tensor. 

In this paper, we propose an iterative method, which solves the following
equation
\begin{eqnarray}
 {E}_r^{(m+1)} = E_r^{*} 
 + \Delta t \mathcal{S}_E(E_r^{(m+1)},
  {F}_r^{(m+1),j}, P_r^{(m+1),jk}, \mathcal{P}_h^{(m)}),\nonumber \\
  {\label{eq:gimp2-1}}\\
 {F}_r^{(m+1),i} = F_r^{*,i} 
 + \Delta t \mathcal{S}_F^i(E_r^{(m+1)},
  {F}_r^{(m+1),j}, P_r^{(m+1),jk}, \mathcal{P}_h^{(m)}),\nonumber \\
  {\label{eq:gimp2-2}}
\end{eqnarray}
where $m=0,1,2...$ indicates the iteration step. 
We take $E_r^{(0)}=E_r^n$, $F_r^{(0),j}=F_r^{n,j}$, and
$\mathcal{P}_h^{(0)}=\mathcal{P}_h^n$ for the initial guess
(also $E_r^*$ and $F_r^{*,j}$ can be
candidates for the initial guess). 
 We note that the hydrodynamic quantity is
explicitly evaluated at $m$-th step
due to the complexity discussed above. 

Next we introduce the following two variables
\begin{eqnarray}
\delta E_r^{(m+1)}&\equiv& E_r^{(m+1)} - E_r^{(m)},\label{eq:defdE}\\
\delta F_r^{(m+1),i}&\equiv& F_r^{(m+1),i} - F_r^{(m),i}.\label{eq:defdF}
\end{eqnarray}
By substituting equations (\ref{eq:defdE}) and (\ref{eq:defdF}) 
into equations (\ref{eq:gimp2-1}) - (\ref{eq:gimp2-2}), 
and taking the first order Taylor series
in $\delta E_r^{(m+1)}$ and $\delta \bmath F_r^{(m+1)}$, we
obtain
\begin{eqnarray}
 \bmath C^{(m)}
 \left(\begin{array}{c}
  \delta E_r^{(m+1)}\\
  \delta F_r^{(m+1),j}\\
  \end{array}\right)=
 \left(\begin{array}{c}
  \Delta t 
   S_E^{(m)}
   +E_r^* - E_r^{(m)} \\
	\Delta t S_F^{(m),i}
	 +F_r^{*,i} - F_r^{(m),i}
	  \end{array}\right),\label{eq:imp2_mat}
\end{eqnarray}
where $S_E^{(m)}=S_E[E_r^{(m)}, F_r^{(m),j}, P_r^{(m),jk},
   \mathcal{P}^{(m)}]$ and $S_F^{(m),i}=
   S_F^{i}[E_r^{(m)}, F_r^{(m),j}, P_r^{(m),jk},
	 \mathcal{P}^{(m)}]$.
Here $\bmath{C}$ is the $4\times 4$ matrix given by
\begin{eqnarray}
\bmath C \equiv \bmath 1 - \Delta t
  \left(\begin{array}{cc}
   \frac{\partial S_E}{\partial E_r}
    +\frac{\partial P^{kl}}{\partial E_r}
    \frac{\partial S_E}{\partial P^{kl}}, 
    \frac{\partial S_E}{\partial F_r^j}
    +\frac{\partial P^{kl}}{\partial F_r^j}
    \frac{\partial S_E}{\partial P^{kl}} \\
	 \frac{\partial S_F^i}{\partial E_r}
	  +\frac{\partial P^{kl}}{\partial E_r}
	  \frac{\partial S_F^i}{\partial P^{kl}}, 
	  \frac{\partial S_F^i}{\partial F_r^j}
	  +\frac{\partial P^{kl}}{\partial F_r^j}
	  \frac{\partial S_F^i}{\partial P^{kl}}
	\end{array}\right),\label{eq:imp2_matC}
\end{eqnarray}
where
\begin{eqnarray}
 \frac{\partial S_E}{\partial E_r}&=&
  -\kappa \rho \gamma +
  \sigma_s \rho \gamma u^2,\label{imp2:dSEdE}\\
 \frac{\partial S_E}{\partial F^j_r}&=&
  \kappa \rho u^j
  -\sigma_s \rho (\gamma^2 +u^2) u^j,\\
 \frac{\partial S_E}{\partial P_r^{kl}}&=&
  \sigma_s \rho \gamma u_k u_l,\\
 \frac{\partial S_F^i}{\partial E_r}&=&
  \sigma_s \rho \gamma^2 u^i,\\
 \frac{\partial S_F^i}{\partial F_r^j}&=&
  -\kappa \rho \gamma \delta^i_j
  -\sigma_s \rho \gamma \left(\delta^i_j + 2 u^i u_j\right),\\
 \frac{\partial S_F^i}{\partial P_r^{kl}}&=&
  \kappa \rho \delta^i_k u_l
  +\sigma_s\rho \left(\delta^i_k u_l+u^i u_k u_l\right).\label{imp2:dSFdP}
\end{eqnarray}

Here we use equations (\ref{geq:G0}), (\ref{geq:Gi}), and (\ref{def:source}).
Also $(\partial P_r^{ij})/(\partial E_r)$ and $(\partial
P_r^{ij})/(\partial F_r^k)$ are required to complete matrix elements. 
These quantities depend on the closure relation given in
equation (\ref{geq:closure}). 
We do not specify the closure relation here, but the explicit form of
these quantities is shown in \S~\ref{closure} and appendix \ref{apPr}.

When $P^{ij}_r$ is the linear function of $E_r$ and $F_r^{i}$ (e.g., the
Eddington approximation), equations (\ref{eq:gimp2-1}) - (\ref{eq:gimp2-2})
reduce to
\begin{eqnarray}
 \bmath C^{(m)}
 \left(\begin{array}{c}
  E_r^{(m+1)}\\
  F_r^{(m+1),i}\\
  \end{array}\right)=
 \left[\begin{array}{c}
    E^n_r + S_E^{(m)}(E_r=0, F_r^i=0)\Delta t\\
  	F^{n,i}_r +S_F^{(m),i}(E_r=0, F_r^i=0)\Delta t
       \end{array}\right].\nonumber \\
 \label{eq:imp2_matlin}
\end{eqnarray}

By inverting the $4\times 4$ matrix $\bmath C$ in equation
(\ref{eq:imp2_mat}) or (\ref{eq:imp2_matlin}), we obtain the radiation
energy and flux at $(m+1)$-th iteration step from equations
(\ref{eq:defdE}) and (\ref{eq:defdF}). 
In general, the matrix inversion is time-consuming and sometimes unstable. The
matrix $C$ is however, only $4 \times 4$ matrix so that we can invert it
analytically. We also tried inverting it using LU-decomposition method.
We obtain the inverse matric $C^{-1}$ stably in both scheme, but the 
analytical method is faster than the LU-decomposition method. Thus we decide to use
analytical expression of $C^{-1}$.

Next, we calculate the primitive variables $\mathcal{P}^{(m+1)}$ from updated
conservative variables $\mathcal{U}^{(m+1)}$. 
As pointed in \S~\ref{Method},  $E_r$ and $F_r^{k}$ are the both
conserved and primitive variables. 
Thus, the recovery step is unnecessary for the radiation
field. We need to compute $\mathcal{P}_h^{(m+1)}$ from
$\mathcal{U}^{(m+1)}$.

Since the total energy $E_t^{n+1}$ and the momentum $m_t^{n+1,k}$ are
already determined, the energy and momentum of fluids
($E_\mathrm{HD}^{(m+1)}$ and $m^{(m+1),k}_\mathrm{HD}$) can be calculated as
\begin{eqnarray}
 E_\mathrm{HD}^{(m+1)} = E_t^{n+1} - E_r^{(m+1)}, \label{eq:Emhd}\\
 m^{(m+1),k}_\mathrm{HD} = m_t^{n+1,k} - F^{(m+1),k}_r. \label{eq:Fmhd}
\end{eqnarray}
Then, three unknown variables $\rho^{(m+1)}, u^{(m+1),k}, p_g^{(m+1)}$ are
computed from 
$D^{n}$, $m^{(m+1),k}_\mathrm{HD}$, and $E^{(m+1)}_\mathrm{HD}$.
Thus, the recovery step in RRHD is the same
as that in relativistic HD. 
We adopt the recovery method developed by \cite{2009ApJ...696.1385Z}
for solving a quartic equation. We
briefly show the method. In the following discussion, we drop the
superscripts $n$ and $(m+1)$ for simplicity. 

The gas density $D$, momentum $m^i_\mathrm{HD}$, and energy
$E_\mathrm{HD}$ are related to the primitive variables $\rho, u^i, p_g$
as 
\begin{eqnarray}
 D=\rho \gamma,\label{eq:Ddef}\\
 m^i_\mathrm{HD} = (\rho + \Gamma_1p_g)\gamma u^i,\label{eq:mdef}\\
  E_\mathrm{HD}  = (\rho + \Gamma_1 p_g)\gamma^2,\label{eq:Edef}
\end{eqnarray}
where $\Gamma_1 \equiv \Gamma/(\Gamma-1)$. 
From equation (\ref{eq:mdef}), we obtain
\begin{equation}
 p_g = \frac{1}{\gamma\Gamma_1}\left(\frac{|m_\mathrm{HD}|}{\sqrt{\gamma^2-1}} - D\right),\label{eq:pgsolv}
\end{equation}
where and $m_\mathrm{HD}=\sqrt{\eta_{ij} m_\mathrm{HD}^i m_\mathrm{HD}^j}$. 
By substituting equation (\ref{eq:pgsolv}) into (\ref{eq:Edef}), we obtain
a quartic equation on $u=\sqrt {u_i u^i}$
\begin{eqnarray}
 f(u)\equiv \Gamma_1^2\left(E_\mathrm{HD}^2 - m_\mathrm{HD}^2 \right) u^4
- 2 \Gamma_1 m_\mathrm{HD} D u^3\nonumber\\
+ \left[\Gamma_1^2 E_\mathrm{HD}^2 - D^2 - 2 \Gamma_1(\Gamma_1-1)|
  m_\mathrm{HD}|^2\right]u^2\nonumber\\
 -2(\Gamma_1-1)D| m_\mathrm{HD}| u - (\Gamma_1-1)^2 m_\mathrm{HD}^2=0.
  \label{eq:primquart}
\end{eqnarray}
By solving above quartic equation, $p_g$, $\rho$ and $u^i$ are computed
from equation (\ref{eq:pgsolv}), (\ref{eq:Ddef}), and (\ref{eq:mdef}).
When we solve equation (\ref{eq:primquart}), we first adopt a Brown
method \citep{2003TJSIAM}, which gives 
analytical solutions of a quartic equation. If reasonable solutions are
not obtained, we solve equation (\ref{eq:primquart}) using
Newton-Raphson method with an accuracy of $f(u) \le 10^{-8}$. 
It is, however, noted that numerical solution converges without switching to
Newton-Raphson method in the test problems shown in \S~\ref{test}.
Also we note that \cite{2009ApJ...696.1385Z} showed that the Brown
method can be applicable to obtain solutions with a wide range of
parameters.

When all the primitive variables are recovered, we obtain solutions
$\mathcal{P}^{(m+1)}$. These solutions would satisfy equation
(\ref{eq:imp2_matlin}). But we note that while $E_r, F_r^i$ and $P_r^{ij}$
are evaluated at $(m+1)$-th time step, we evaluate
$\mathcal{P}_h$ at $(m)$-th time step when we solve equation
(\ref{eq:imp2_matlin}). The evaluation of $\mathcal{P}_h$
at $(m)$-th iteration step might be problematic when the density or
temperature jump is large (e.g., at the shock front).
Thus, we again solve equation (\ref{eq:imp2_matlin}) using updated
primitive variables $\mathcal{P}_h^(m+1)$ and check the difference
between two successive variables,
$|\delta E_r^{(m+1)}|/E_r^{(m+1)}$, 
$|\delta F_r^{(m+1),i}/F_r^{(m+1),i}|$, and 
$|\mathcal{P}_h^{(m+1)}-\mathcal{P}_h^{(m)}|/|\mathcal{P}^{(m+1)}_h|$. 
When these quantities is larger than a specified value (typically $\sim
10^{-6}$), the source term is integrated again using updated primitive
variables. This process is continued until two successive variables fall
below a specified tolerance \citep{2009MNRAS.394.1727P}. 
Similar iteration method is also adopted in relativistic resistive
magnetohydrodynamics \citep{2009MNRAS.394.1727P}. Since $\mathcal{P}_h$
is evaluated at the current iteration step, solutions obtained by solving
equation (\ref{eq:imp2_matlin}) might not be converged when the shock
appears. But as we will see in \S~\ref{test}, we obtain solutions correctly even when the
cooling time is much shorter than $\Delta t$ (see,
\$~\ref{radheatcool}), or a shock appears (see, \S~\ref{RHD}).

Now we summarize our newly developed scheme for solving RRHD equations.
\begin{enumerate}
 \item Calculate primitive variables at the cell surface
       $\mathcal{P}_{i\pm 1/2, S}$ from $\mathcal{P}_i$. 
 \item Compute the numerical flux $f_{i\pm 1/2}$ using approximate
       Riemann solvers (e.g., HLL scheme). 
 \item Integrate the hyperbolic term from numerical flux $f_{i\pm 1/2}$ (in
       equation \ref{geq:fluxterm}). Then we obtain the intermediate states of
       conservative variables $\mathcal{U}^*$.
 \item Compute matrix elements of $\bmath C$ given in equation
       (\ref{eq:imp2_matC}).
       Then, calculate $E_r^{(m+1)}$ and $F_r^{(m+1),i}$ by inverting the
       $4\times 4$ matrix.
 \item Calculate primitive variables $\mathcal{P}^{(m+1)}$ from
       $E_\mathrm{HD}^{(m+1)}$, $m_\mathrm{HD}^{(m+1),k}$ and $D^{n+1}$. 
 \item When the difference between two successive values does not fall
       below a specified tolerance, repeat from step 4. 
       The updated primitive variables of fluids $\mathcal{P}_h^{(m+1)}$ are
       used to evaluate matrix $\bmath C$.
\end{enumerate}

\subsection{Closure Relation}\label{closure}
In the above discussion, we do not specify the closure relation, but 
take the general form given in equation (\ref{geq:closure}).
When we compute the numerical flux using the HLL scheme, the wave speeds of
the fastest mode $\lambda_\pm$ are needed, which depend on the closure
relation.
Another term depending on the closure relation is $P_r^{ij}$, which
appears in numerical flux (\ref{geq:Flux}).
Also we need to evaluate its derivatives, $\partial P^{ij}/\partial E_r$
and $\partial P^{ij}/\partial F_r^{k}$ to compute the matrix $\bmath C$.
In this subsection, we show how to obtain these quantities by specifying
the closure relation.

Many kinds of closure relation are proposed by authors as discussed in
\S~\ref{intro}. As the first step of developing the RRHD code,
we hereafter restrict our discussion to the closure relation being provided when
we assume the Eddington approximation. 
Then, the closure relation is described by
\begin{equation}
 P'^{ij}_r = \frac{\delta^{ij}}{3}E'_r, \label{eq:Eddington}
\end{equation}
which is valid when the radiation is well coupled with the matter so
that the radiation field is isotropic in the comoving frame. This
corresponds to the Eddington factor being $1/3$.

First, we show the wave speed of the fastest mode.
By assuming the relation given in equation (\ref{eq:Eddington}), the
characteristic wave velocity of fastest mode in the comoving frame is
$1/\sqrt{3}$, which is equivalent to the sound speed in the relativistic
regime. Thus the maximum wave velocity is always $1/\sqrt{3}$ in the
radiation hydrodynamics with the Eddington approximation.
The wave velocity in the observer frame $\lambda_-$
and $\lambda_+$ is obtained
by boosting $\pm 1/\sqrt{3}$ with the fluid velocity.

Next, we show how to obtain radiation pressure $P_r^{ij}$ measured in
the observer frame. By performing Lorentz transformation on the radiation
energy momentum tensor and combining it with equation
(\ref{eq:Eddington}), the radiation stress tensor in the observer frame
obeys the following equation
\begin{eqnarray}
 P_r^{ij}
&+&\left[-\frac{\delta^{ij}}{3}+\frac{u^i u^j }{(1+\gamma)^2} \right] 
u_k u_mP_r^{km} \nonumber \\
 &+& \frac{1}{1+\gamma} \left(u^i u_k P_r^{jk} + u^j u_k P_r^{ik}\right)
   = R^{ij},\label{eq:Edclos}
\end{eqnarray}
where
\begin{eqnarray}
 R^{ij}&=&
  \frac{\delta^{ij}}{3}\left(\gamma^2 E_r - 2 \gamma u_k F_r^k \right)
  -u^i u^j E_r \nonumber \\
 &+& (u^i F_r^j + u^j F_r^i)
  +  \frac{2}{1+\gamma}u^i u^j u_k F_r^k,\label{eq:EdclosR}
 \end{eqnarray}
\citep[e.g.,][]{2008bhad.book.....K}. 
Since the radiation stress $P_r^{ij}$ is a $3\times 3$ symmetric matrix, we
need to solve 6th order linear equations to compute $P_r^{ij}$ as
\begin{equation}
 \bmath A(u)p = r,\label{eq:Plinear}
\end{equation}
where $p^T=(P_r^{11}, P_r^{22}, P_r^{33},P_r^{12},P_r^{13},P_r^{23})$,
$r^T=(R^{11}, R^{22}, R^{33},R^{12},R^{13},R^{23})$ and $\bmath A=\bmath
A(u)$ is the
$6\times 6$ matrix. Since $\bmath A$ is a function of velocity, the derivatives
of $P_r^{ij}$ are described by
\begin{equation}
 \bmath A(u)\frac{\partial p}{\partial E_r} = \frac{\partial r}{\partial E_r},~
  \bmath A(u)\frac{\partial p}{\partial F^k_r} = \frac{\partial
  r}{\partial F^k_r}.   
\end{equation}
Thus, $P_r^{ij}$, $\partial P_r^{ij}/\partial E_r$
and $\partial P_r^{ij}/\partial F_r^{k}$ are computed by inverting matrix
$\bmath A$. 
Since $\bmath A$ is a $6 \times 6$ matrix, it is difficult to obtain
$\bmath A^{-1}$ using analytical expression. Thus, we use LU-decomposition
method to invert matrix $\bmath A$. 
Explicit forms of
$\bmath A, (\partial r)/(\partial E_r)$, and $(\partial r)/(\partial F_r^k)$
are shown in appendix \ref{apPr}.

\begin{figure}
 \begin{center}
  \includegraphics[width=8cm]{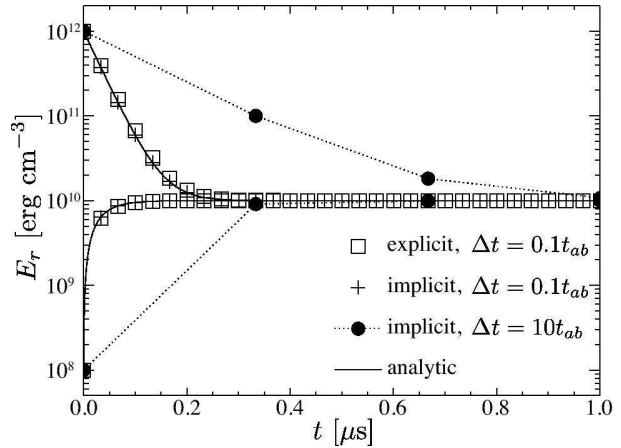}
  \caption{Thermal evolution of radiation energy $E_r$. Crosses and
  squares denote results of explicit and implicit schemes, while solid
  curves do analytical solutions. The filled circles with dotted curves show
  the results of the implicit scheme with the time step of $\Delta t =
  10t_{ab}$.}
  \label{fig:radheatcool}
 \end{center}
\end{figure}

\section{Test Problems}\label{test}
In this section, we show results of some numerical tests for our RRHD
code. This section consists of numerical tests for the radiation field
(\S~\ref{RD}) and relativistic radiation hydrodynamics (\S~\ref{RHD}).
We assume that the closure 
relation is given by equation (\ref{eq:Eddington}) so that the wave
speed of radiation field is $1/\sqrt{3}$.

\begin{figure*}
 \begin{tabular}{cc}
 \begin{minipage}{0.49\hsize}
   \includegraphics[width=8cm]{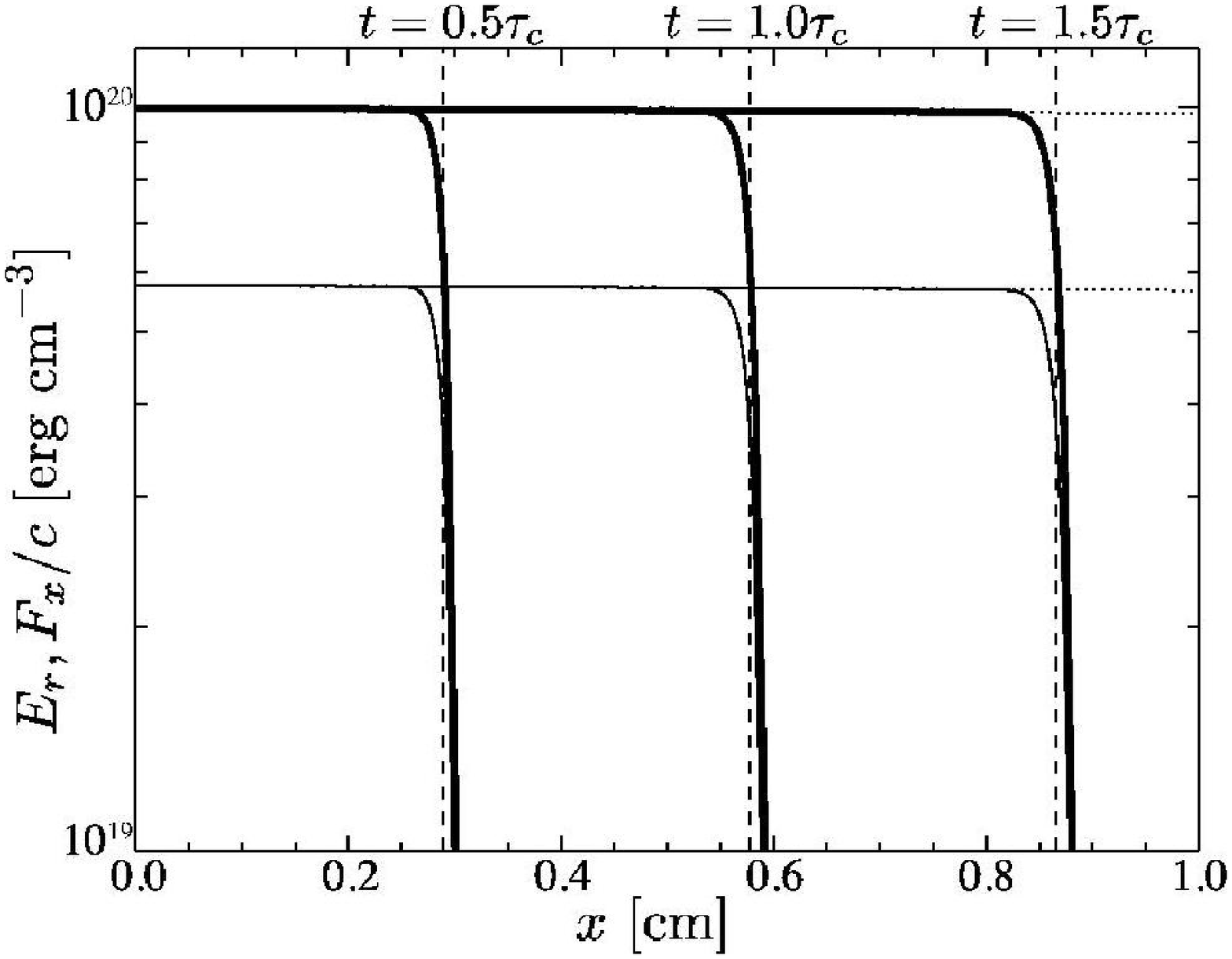}
   \caption{Time evolution of $E_r$ (thick solid) and $F_r$ (solid). The
  optical depth of the system size is $0.01$. Dashed curves denote 
  the light head $l_c = c t/\sqrt{3}$ at $t=0.5, 1.0, 1.5\tau_c$,
  where $\tau_c = L/c$. Dotted curves show analytical solutions assuming
  steady state given in equation (\ref{eq:radflowst}).}
   \label{fig:radtrans1}
 \end{minipage}
  \hspace{1mm}
  \begin{minipage}{0.49\hsize}
   \includegraphics[width=8cm]{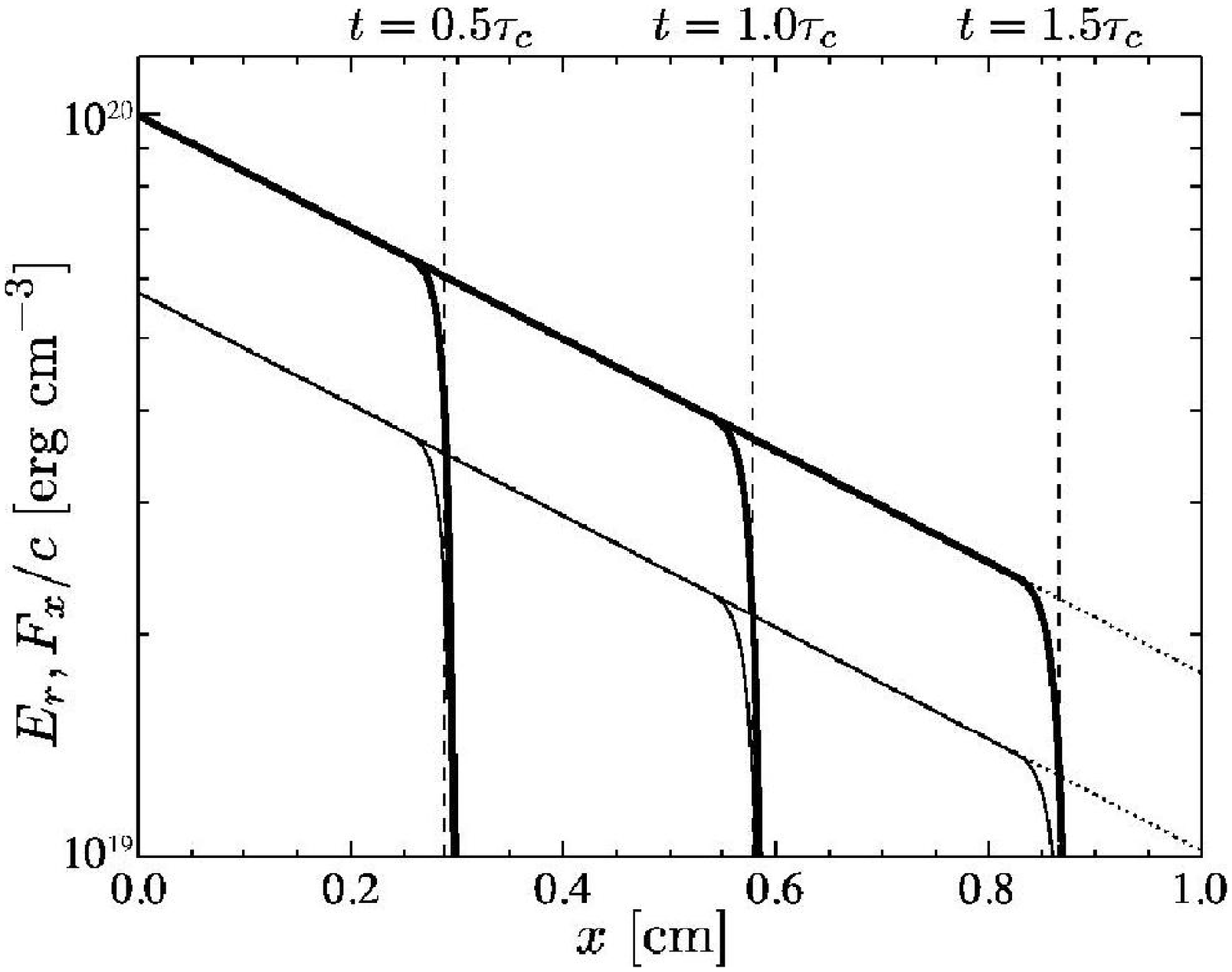}
   \caption{Time evolution of $E_r$ (thick solid) and $F_r$ (solid). The
  optical depth of the system size is unity. Dashed curves denote 
  the light head $l_c = c t/\sqrt{3}$ at $t=0.5, 1.0, 1.5\tau_c$,
  where $\tau_c = L/c$. Dotted curves show analytical solutions assuming
  steady state given in equation (\ref{eq:radflowst}).}
   \label{fig:radtrans2}
  \end{minipage}
 \end{tabular}
\end{figure*}
\begin{figure*}
 \begin{tabular}{cc}
 \begin{minipage}{0.49\hsize}
   \includegraphics[width=8cm]{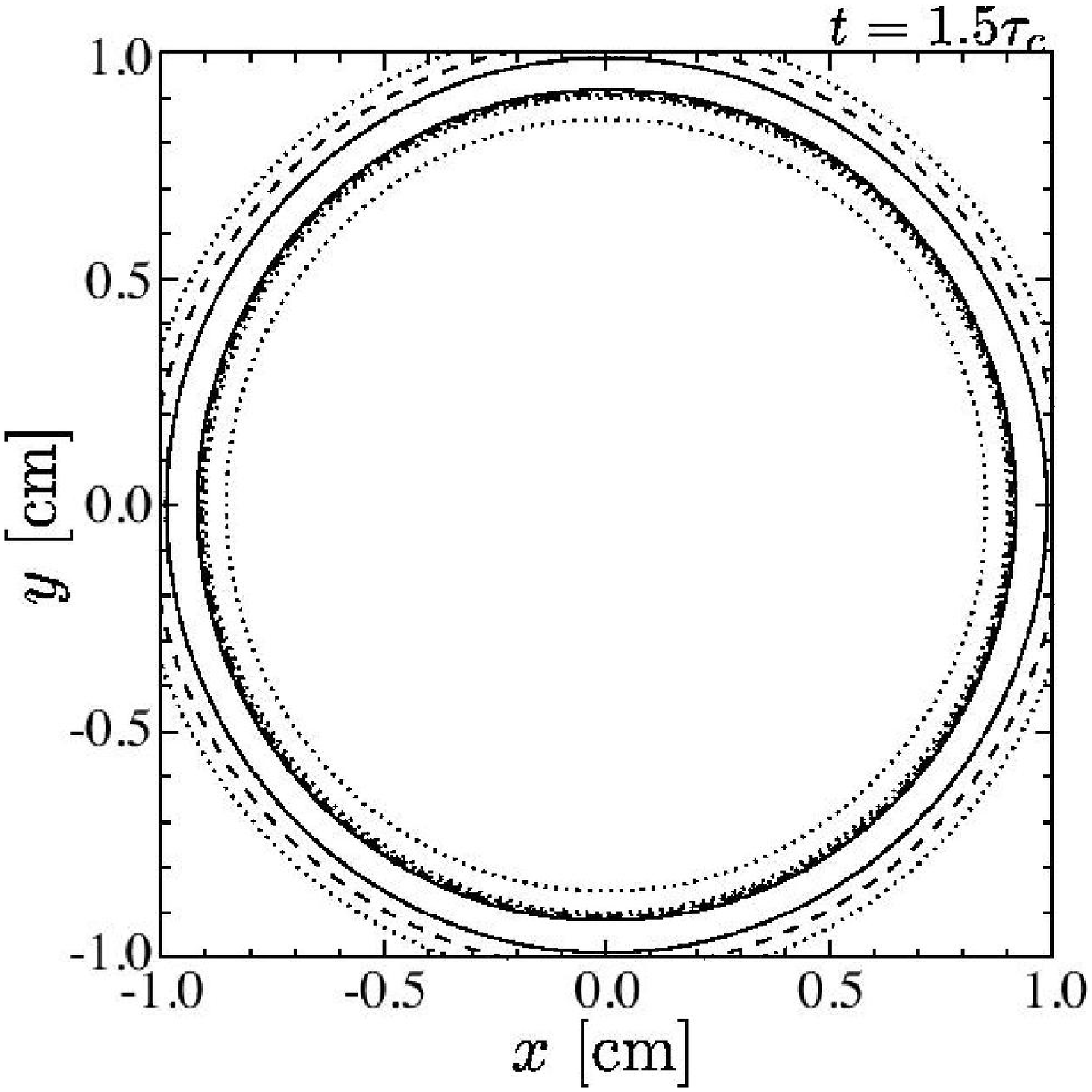}
  \caption{Snapshot of $E_r$ at $t=1.5\tau_c$. The solid, dashed and dotted
  curves denote contours at $10\%, 0.1\%, 10^{-3}\%$ of its maximum.}
   \label{fig:radtrans3}
 \end{minipage}
  \hspace{1mm}
  \begin{minipage}{0.49\hsize}
   \includegraphics[width=8cm]{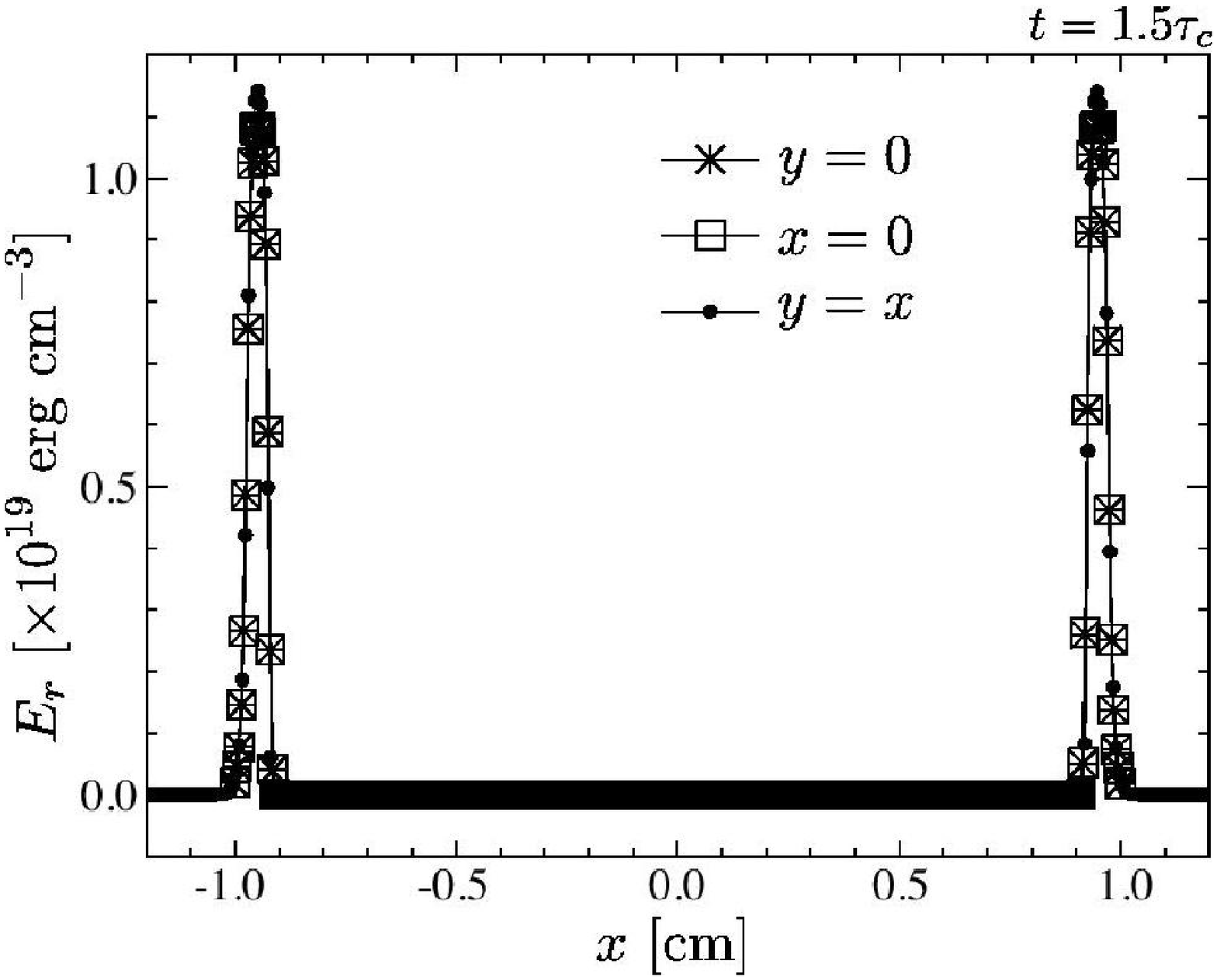}
   \caption{Spatial profiles of $E_r$ at $t=1.5 \tau_c$. Asterisks,
   squares, and filled circles indicate results along $y=0$, $x=0$,
   and $y=x$, respectively.}
   \label{fig:radtrans4}
  \end{minipage}
 \end{tabular}
\end{figure*}
\subsection{Numerical tests for Radiation Field}\label{RD}
In this section, we show results of numerical tests for solving
radiation fields given in equations (\ref{geq:Tradcons}).
We assume that the fluid is static and uniform for simplicity. We
recover the light speed as $c$ in this subsection. 
Equations are solved with 2nd order accuracy in space.

\subsubsection{Radiative heating and cooling}\label{radheatcool}
This test has been proposed by \cite{2001ApJS..135...95T}.
We evaluate the validity for the integration of the source
term appeared in equation (\ref{geq:sourceterm}), which is implicitly and
iteratively integrated in our numerical scheme. For this purpose, we start
from a
static and 1 zone fluid (i.e., number of grid points $N_x = 1$), which
is initially not in
the thermal equilibrium with the radiation. 
From these assumptions, the radiation field obeys the following equation
\begin{equation}
 \frac{d E_r}{d t} = \rho \kappa 
  \left(4\pi \mathrm{B}-c  E_r\right),
\end{equation}
which is analytically integrated by assuming $\rho, \kappa$ and
$\mathrm{B}$ are constant, and we obtain
\begin{equation}
 E_r(t) = \frac{4\pi}{c} \mathrm{B}
  - \left(\frac{4\pi}{c} \mathrm{B}-E_0\right)e^{-\rho \kappa  c t},
\label{eq:radheatcool}
\end{equation}
where $E_0$ is the radiation energy density at $t=0$.
The radiation field approaches that in the Local Thermal Equilibrium
(LTE, $E_r = 4\pi \mathrm{B}/c$). 

The mass density is set to $\rho = 0.025~\mathrm{g~cm^{-3}}$ and the
opacity is $\kappa = 0.04~\mathrm{cm^{2}~g^{-1}}$, so that the
corresponding absorption time scale is $t_{ab}\equiv 1/(\rho\kappa
c)=3.3\times 10^{-8}~\mathrm{s}$.
We examine two models of the initial radiation energy density,
$E_r=10^{2} E_\mathrm{LTE}$ and $10^{-2} E_\mathrm{LTE}$, where
$E_\mathrm{LTE}\equiv a_R T^4=10^{10}~\mathrm{erg~cm^{-3}}$. The
specific heat ratio and the mean
molecular weight are $\Gamma=5/3$ and $\mu=1.0$. 
\begin{deluxetable*}{lccccccccccccc}
\tabletypesize{\scriptsize}
\tablecaption{List of Simulation Runs}
\tablewidth{0pt}
\tablehead{
\colhead{model} & \colhead{$\kappa$} &\colhead{$\Gamma$}
 &\colhead{state} & \colhead{$\rho$} &
 \colhead{$p_g$} & \colhead{$u^x$} & \colhead{$u^y$}
 & \colhead{$u^z$} & \colhead{$E_r'$}
}
\startdata
 RHDST1       &0.4 &$\frac{5}{3}$ & 
 L & 1.0      & $3.0\times 10^{-5}$ & 0.015 & 0    & 0  & $1.0\times 10^{-8}$\\
   \          &\ &\ & 
 R & 2.4      & $1.61\times 10^{-4}$ & $6.25\times 10^{-3}$ & 0 & 0 & $2.50\times 10^{-7}$\\
 RHDST2       &0.2 &$\frac{5}{3}$ & 
 L & 1.0      & $4.0\times 10^{-3}$ & 0.25 & 0    & 0  & $2.0\times 10^{-5}$\\
   \          &\ &\ & 
 R & 3.11      & $0.04512$ & $0.0804$ & 0    & 0& $3.46\times 10^{-3}$\\
 RHDST3       &0.3 &2& 
 L & 1.0      & 60 & 10 & 0    & 0     & 2\\
   \          &\ &\ & 
 R & 8      & $2.34\times 10^{3}$ & 1.25 & 0  & 0& $1.13\times 10^{3}$\\
 RHDST4       &0.08 &$\frac{5}{3}$& 
 L & 1.0      & $6.0\times 10^{-3}$ & 0.69 & 0    & 0   & 0.18\\
   \          &\ &\ & 
 R & 3.65      & $3.59\times 10^{-2}$ & 0.189 & 0    & 0& $1.30$\\
\enddata
\tablecomments{Parameter sets of numerical tests. Scattering coefficient
 $\sigma_s$ is taken to be zero in all models.}
\label{tab:testparam}
\end{deluxetable*}

Figure~\ref{fig:radheatcool} shows the time evolution of $E_r$. Squares
and crosses respectively denote results with explicit and
implicit scheme for integrating the source term with the time step of
$\Delta t = 0.1 t_{ab}$, while 
solid curves do analytical solutions obtained from equation
(\ref{eq:radheatcool}). 
The data point for the model with $\Delta t=0.1 t_{ab}$ is reduced for
plotting.
The lower and upper plots
correspond to $E_r(t=0)=10^{-2}E_\mathrm{LTE}$ and $10^{2}E_\mathrm{LTE}$,
respectively. 
We can see that results with both schemes excellently agree
with analytical solutions when the time step is smaller than the absorption
time scale $\Delta t < t_{ab}$. 
When $\Delta t = 10 t_{ab}$, 
solutions with implicit scheme (filled circle) stably reach
the thermal equilibrium state, although
they deviate from the
analytical solutions before being in LTE. 
We note that the explicit scheme does not converge to the analytical
solution when we take such a large time step. 
We also performed simulations with a much larger time step $\Delta t = 10^4 t_{ab}$
and found that solutions converge to analytical solution (not shown in figure).
We also note that a number of iteration step used in the implicit scheme
is less than or equal 2, even if $\Delta t = 10^4 t_{ab}$.
Thus, the implicit scheme has a great advantage when the cooling timescale
is much smaller than the dynamical timescale since we can take the numerical
time step being $t_{ab}<\Delta t < t_{dyn}$.
The computational time of explicit and implicit scheme with an equal number
of time steps is $t_\mathrm{exp} : t_\mathrm{imp}=1 : 1.04$ with $\Delta
t=0.1 t_{ab}$.

\subsubsection{Radiation transport}\label{radtransport}
This test has been performed by \cite{2001ApJS..135...95T}.
When the radiation is injected into uniform matter, it propagates
with the characteristic wave velocity, while it exchanges energies with the
plasma. As a test of this effect, we assume that the fluid is static and
uniform in a simulation box bounded by $x=[0,L]$, where
$L=1~\mathrm{cm}$. 
The radiation is injected from the boundary at $x=0$. 
In the uniform medium, the opacity is set to $\kappa =
0.04~\mathrm{cm^{-2}~g}$. 
The radiation energy is $E_\mathrm{LTE}=10^{10}~\mathrm{erg~cm^{-3}}$
and the thermal energy is determined from the LTE condition.
We examined two models of the mass density $\rho = 0.25$ and
$25~\mathrm{g~\mathrm{cm}^{-3}}$.
The corresponding optical depth is
$\tau = \rho \kappa L=0.01$ and $1$, respectively.
At the boundary $x=0$, the radiation is injected with the energy density
of $E_r = 10^{10}E_\mathrm{LTE}$. The free boundary condition is applied
at $x=L$. 
Other parameters are $\Gamma = 5/3$, and $\mu = 1.0$. 
The Courant-Friedrichs-Lewy (CFL) number is taken to be $0.5$.

Figure~\ref{fig:radtrans1} shows results with $\tau = 0.01$. 
Thick solid curves denote the radiation energy density, and thin solid curves
show the radiation flux at
$t=0.5, 1.0, 1.5\tau_c$ from left to right, where $\tau_c = L/c$. Since the radiation energy
is injected from $x=0$, the wave
front propagates from left to right as time goes on. The dashed curves
denote the position of the light head $l_c=ct/\sqrt{3}$. Since we assume
the Eddington approximation on the closure relation, the wave front
propagates with the velocity $c/\sqrt{3}$. We can see that the wave
front is sharply captured in our simulation code. In FLD approximations,
the radiation field evolves obeying the diffusion equation, so
that the wave front has a smooth profile \cite[see, Fig.~7 in
][]{2001ApJS..135...95T}. In our code, although we apply the Eddington
approximation, the 1st order moment equation is solved. 
Then equations of the radiation field have hyperbolic form, so that the wave
front can be captured using the HLL scheme.

Figure~\ref{fig:radtrans2} shows results of the model of larger density
($\tau=1$). The
radiation propagates with the velocity $c/\sqrt{3}$.
Behind the wave front, the radiation field becomes steady and
its energy exponentially decreases with $x$ (equivalently, $\tau =
\kappa \rho x$) due to the absorption. 
When we assume the steady state and the
radiation energy is much larger than that in LTE (i.e., $E_r \gg 4\pi
\mathrm{B}/c$), equation (\ref{geq:Tradcons}) can be solved and we obtain
\begin{equation}
 E_r = E_0 \exp(-\sqrt{3}\rho \kappa x), ~
  F_r = \frac{cE_0}{\sqrt{3}} \exp(-\sqrt{3}\rho \kappa x),\label{eq:radflowst}
\end{equation}
\citep{1984oup..book.....M}.
Dotted curves in Fig. 
\ref{fig:radtrans1} and \ref{fig:radtrans2} show solutions obtained from
equation (\ref{eq:radflowst}). We can see that
numerical results excellently recover analytical ones.
\begin{figure*}
 \begin{tabular}{cc}
 \begin{minipage}{0.49\hsize}
   \includegraphics[height=9cm]{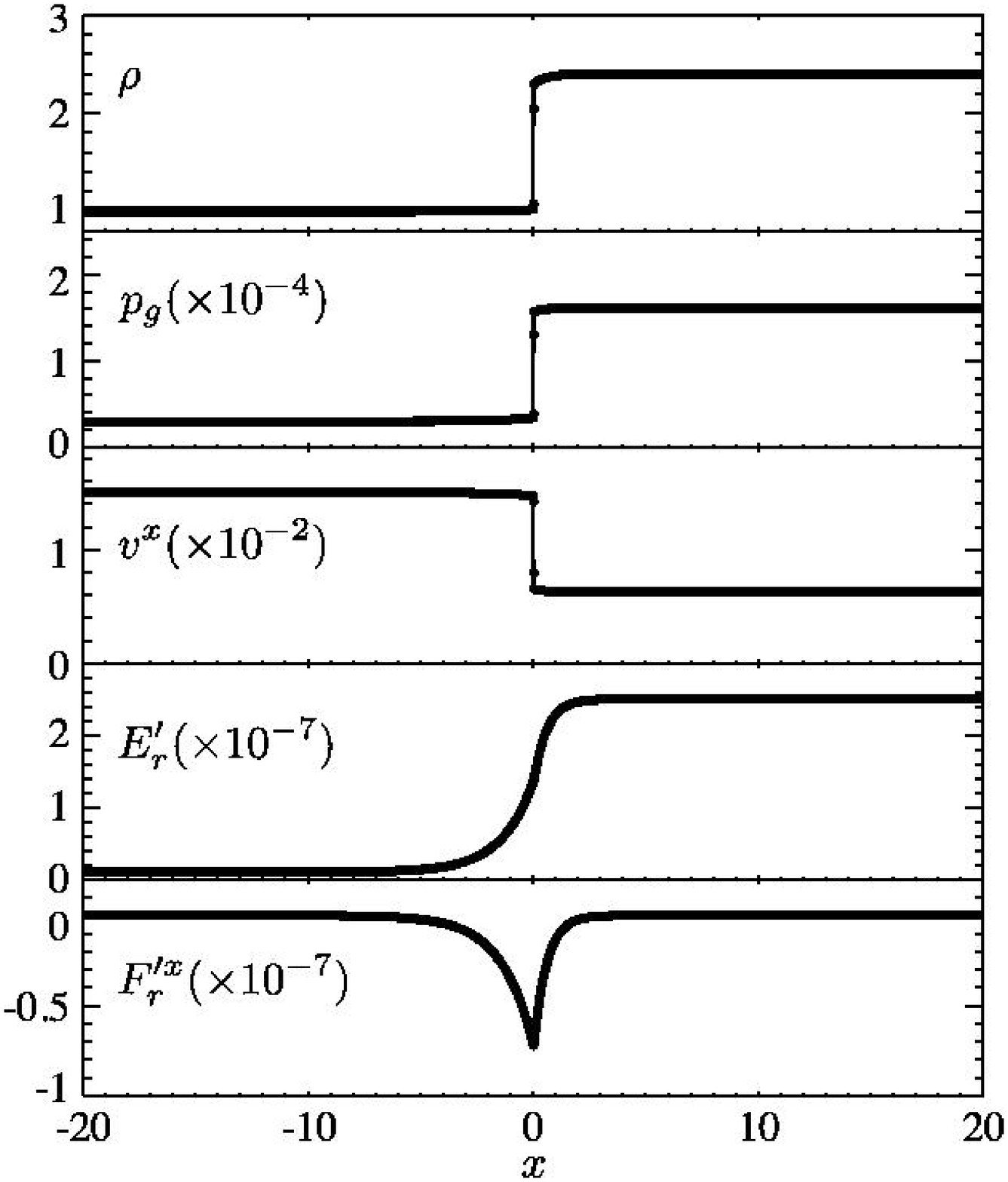}
   \caption{Profiles of $\rho$, $p_g$, $v^x$, $E'_r$ and $F'^x_r$ from
  top to bottom at $t=5000$ for the model RHDST1.
Dots and solid curves denote for numerical and semi-analytical solutions.
}
   \label{fig:farris1}
 \end{minipage}
  \hspace{1mm}
  \begin{minipage}{0.49\hsize}
   \includegraphics[height=9cm]{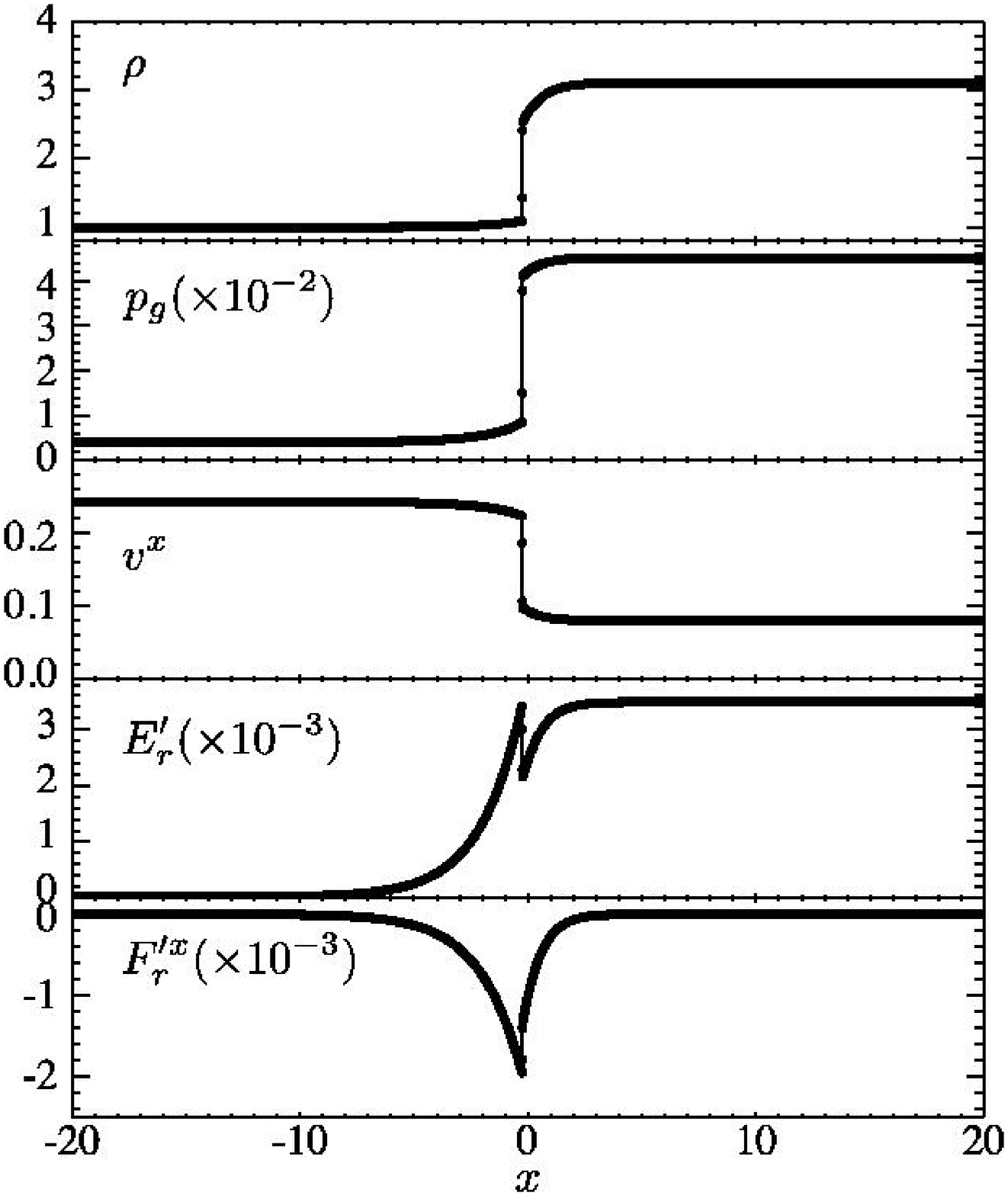}
   \caption{Profiles of $\rho$, $p_g$, $v^x$, $E'_r$ and $F'^x_r$ from
  top to bottom at $t=5000$ for the model RHDST2. 
Dots and solid curves denote for numerical and semi-analytical solutions.}
   \label{fig:farris2}
  \end{minipage}
 \end{tabular}
\end{figure*}

\subsubsection{Radiation transport in 2-dimension}\label{radtransport2}
This test has been shown in \cite{2001ApJS..135...95T} that the radiation
field propagates in optically thin media in $x-y$ plane. 
The simulation domain is $x=[-L,L]$ and $y=[-L, L]$, where
$L=1\mathrm{cm}$. Numerical grid points are $(N_x, N_y)=(400, 400)$. 
The mass density of uniform fluid is $\rho =
0.25~\mathrm{g~cm^{-3}}$. The other quantities are the same as those
in \S~\ref{radtransport}. At the initial state, a larger radiation energy
is given inside $r=\sqrt{x^2+y^2}<0.1\mathrm{~cm}$ and its energy is
$E_r=10^{10}E_\mathrm{LTE}$. The free (Neumann) boundary condition is
applied on each side of the box. The CFL condition is taken as $0.5$. 

Figure \ref{fig:radtrans3} shows contours of radiation energy density
$E_r$ at $t=1.5\tau_c$. 
The solid, dashed, and dotted curves denote contours at $10\%, 0.1\%,
10^{-3}\%$ of its maximum. 
Due to the enhancement of radiation energy around the origin at the
initial state, the radiation field propagates in a circle, forming a
caldera-shaped profile of the radiation energy density. The wave
front propagates with the speed of $c/\sqrt{3}$. 
Figure \ref{fig:radtrans4} shows one dimensional cut of results at
$t=1.5\tau_c$. Asterisks, squares and
filled circles show results along $y=0$, $x=0$, and $y=x$,
respectively. 
We can see that the wave packet has a sharp structure. The radiation energy is
mainly accumulated around the wave front, while it is very small
inside the wave front (caldera floor). Such the caldera structure
is not successfully reproduced when we apply FLD approximation
\citep{2001ApJS..135...95T} since the FLD is formulated based on the
diffusion approximation.
Although the wave speed reduces to $c/\sqrt{3}$, solving the 1st order
moment in radiation field (radiation flux) with Eddington approximation
has an advantage for studying the propagation of the radiation pulse
in optically thin media.

In the problem of $\S~\ref{radtransport}$ and $\S~\ref{radtransport2}$,
the radiation field passes the boundary $x=L$ at
$t=\sqrt{3}\tau_c$. When we adopt the free (Neumann) boundary 
conditions,
most of the radiation energy passes the boundary, while some part of
it is reflected and stays in the simulation box. An amplitude
of the reflected wave $E_\mathrm{ref}$ is smaller than that passing
the boundary $E_\mathrm{pass}$, $\max[E_\mathrm{ref}/E_\mathrm{pass}]\sim
0.8\%$ for the one-dimensional test and $\sim 5\%$ for the two dimensional
test. The waves are mainly reflected at the corner of the simulation box
in the two dimensional simulation (i.e., around $[x,y]=[\pm L, \pm L]$). 
The simple free boundary condition can be applied in the
radiation field with this accuracy.

\begin{figure*}
 \begin{tabular}{cc}
 \begin{minipage}{0.49\hsize}
   \includegraphics[height=9cm]{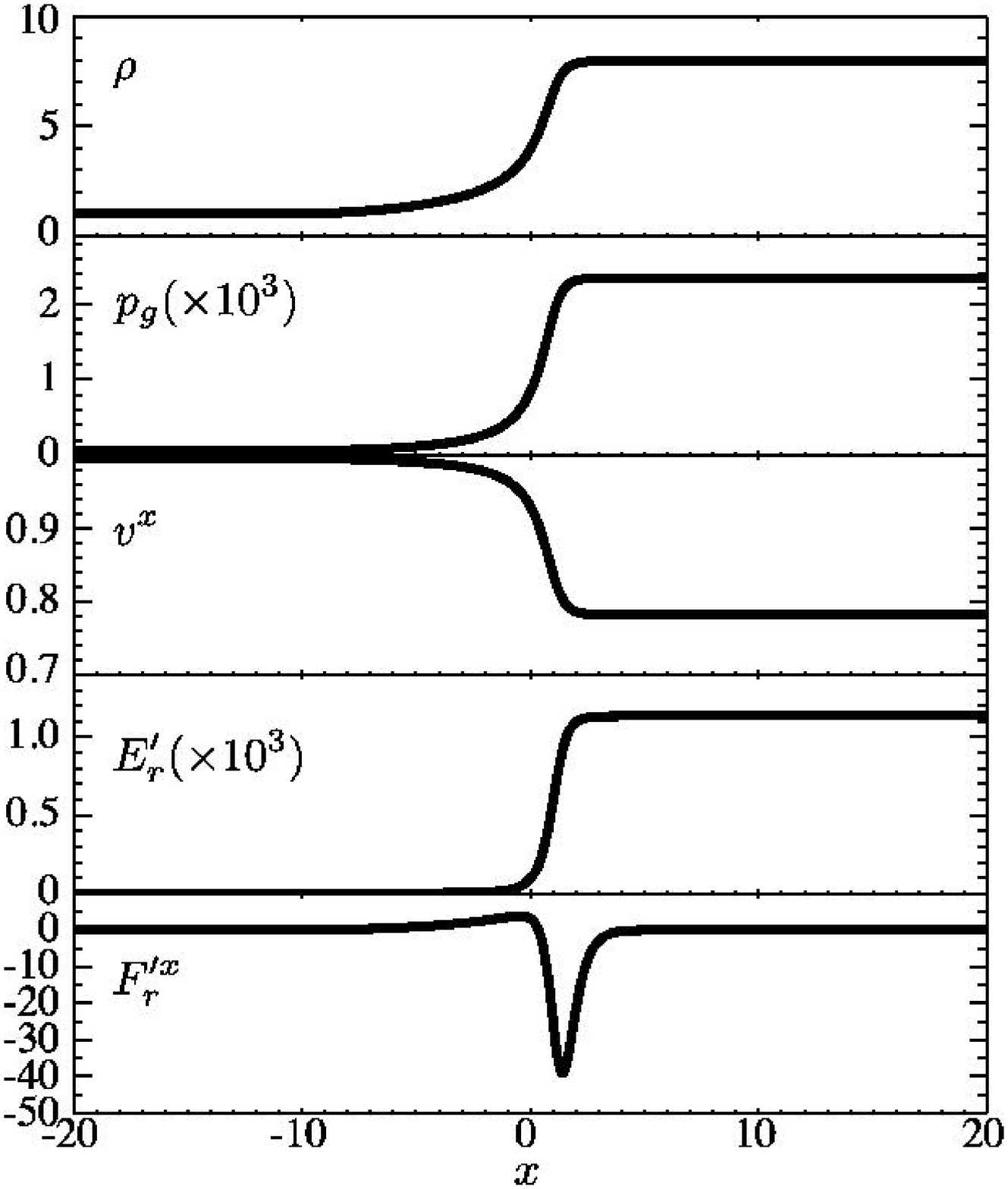}
   \caption{Profiles of $\rho$, $p_g$, $v^x$, $E'_r$ and $F'^x_r$ from
  top to bottom at $t=5000$ for the model RHDST3. Dots and solid curves
  denote for numerical and semi-analytical solutions.}
   \label{fig:farris3}
 \end{minipage}
  \hspace{1mm}
  \begin{minipage}{0.49\hsize}
   \includegraphics[height=9cm]{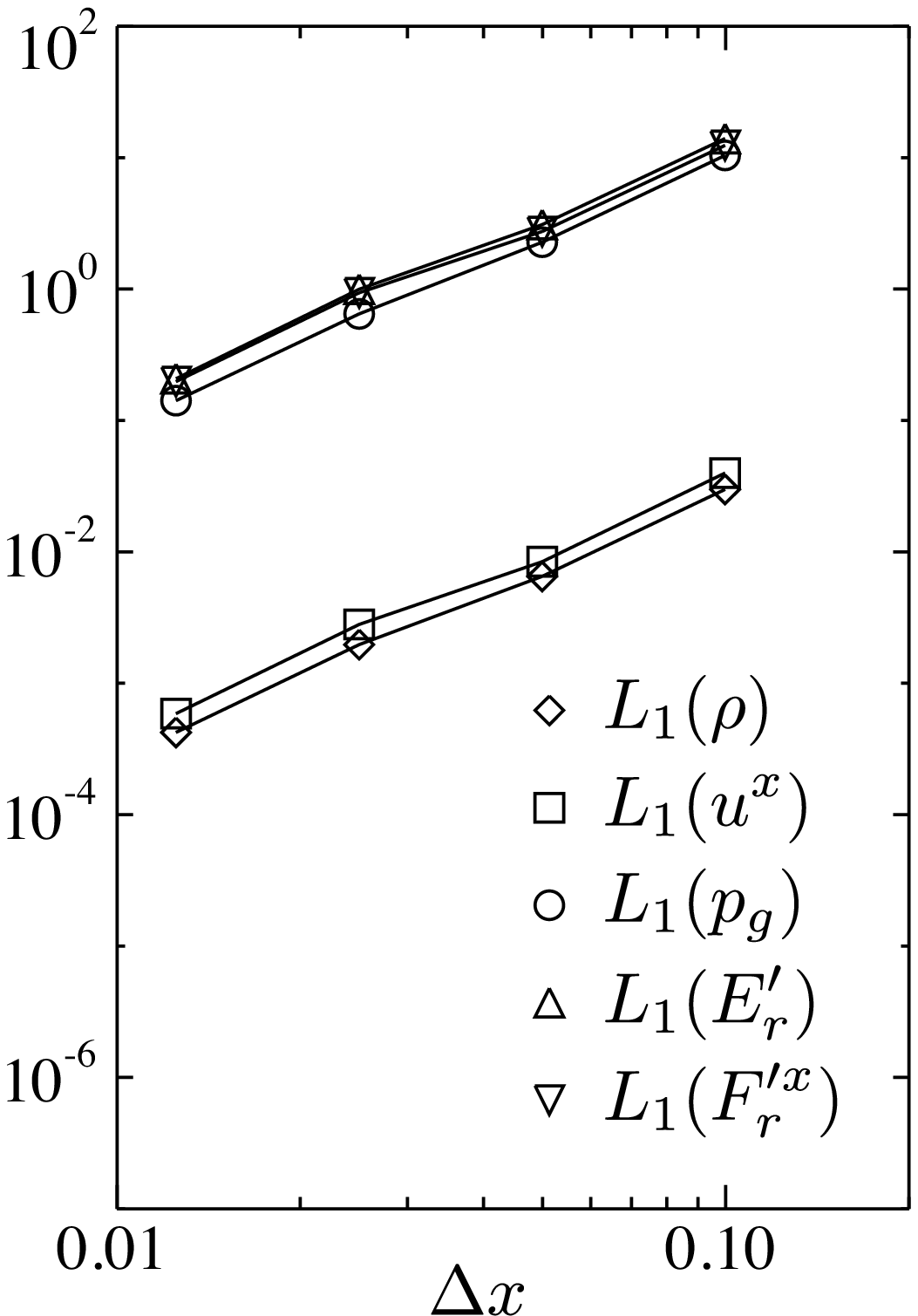}
   \caption{$L$-1 norms of $\rho$, $p_g$, $u^x$, $E'_r$ and $F'^x_r$ for
   the model RHDST3.}
   \label{fig:farris3-L1}
  \end{minipage}
 \end{tabular}
\end{figure*}

\subsection{Numerical Test for Relativistic Radiation Hydrodynamics}\label{RHD}
Next, we consider the coupling between the radiation and
matter fields. We solve RRHD equations with a second order accurate
scheme. 

Test problems for radiative shock shown in this subsection are developed
by \cite{2008PhRvD..78b4023F} who proposed and solved four shock tube
test problems. 
Initial conditions of these problems are listed in
Table~\ref{tab:testparam}. The initial state has a discontinuity at
$x=0$ and the system is in LTE. 
Such an initial discontinuity breaks up generating waves. 
When the waves pass away from the simulation box, the system approaches
the steady state. Then we compare our numerical results with
semi-analytical solutions
after the system reaches a steady state.
In \cite{2008PhRvD..78b4023F}, the initial condition is constructed by
boosting semi-analytical solutions, so that the shock moves with an
appropriate velocity. In our test, the shock is assumed to rest
around $x=0$ (shock rest frame).
Such the problem can be more stringent tests for
our code to maintain the stationarity \citep{2011MNRAS.tmp.1386Z}.

The simulation box is bounded by $x=[-L, L]$, where $L=20$ in the
normalized unit,  and a number of numerical
grid points is $N_x=3200$. The free boundary condition is applied at
$x=\pm L$. 
We give the physical quantities at the left and
right state of $\rho, p, u, E_r'$ and flux is assumed to be zero. 
Following \cite{2008PhRvD..78b4023F}, the Stefan-Boltzmann constant is
normalized such that $4\pi a_R=E'_{r,L}/T_L^4 =
E'_{r,R}/T_R^4$, where the subscripts $L$ and $R$ denote the quantities
at the left ($x\leq 0$) and right ($x>0$). 

We note that although we adopt an iteration method to integrate
 stiff source term (equation (\ref{eq:imp2_mat})), solutions in the
 following tests converge within the relative error of $10^{-6}$ without iterations. 

\begin{figure*}
 \begin{tabular}{cc}
 \begin{minipage}{0.49\hsize}
   \includegraphics[height=9cm]{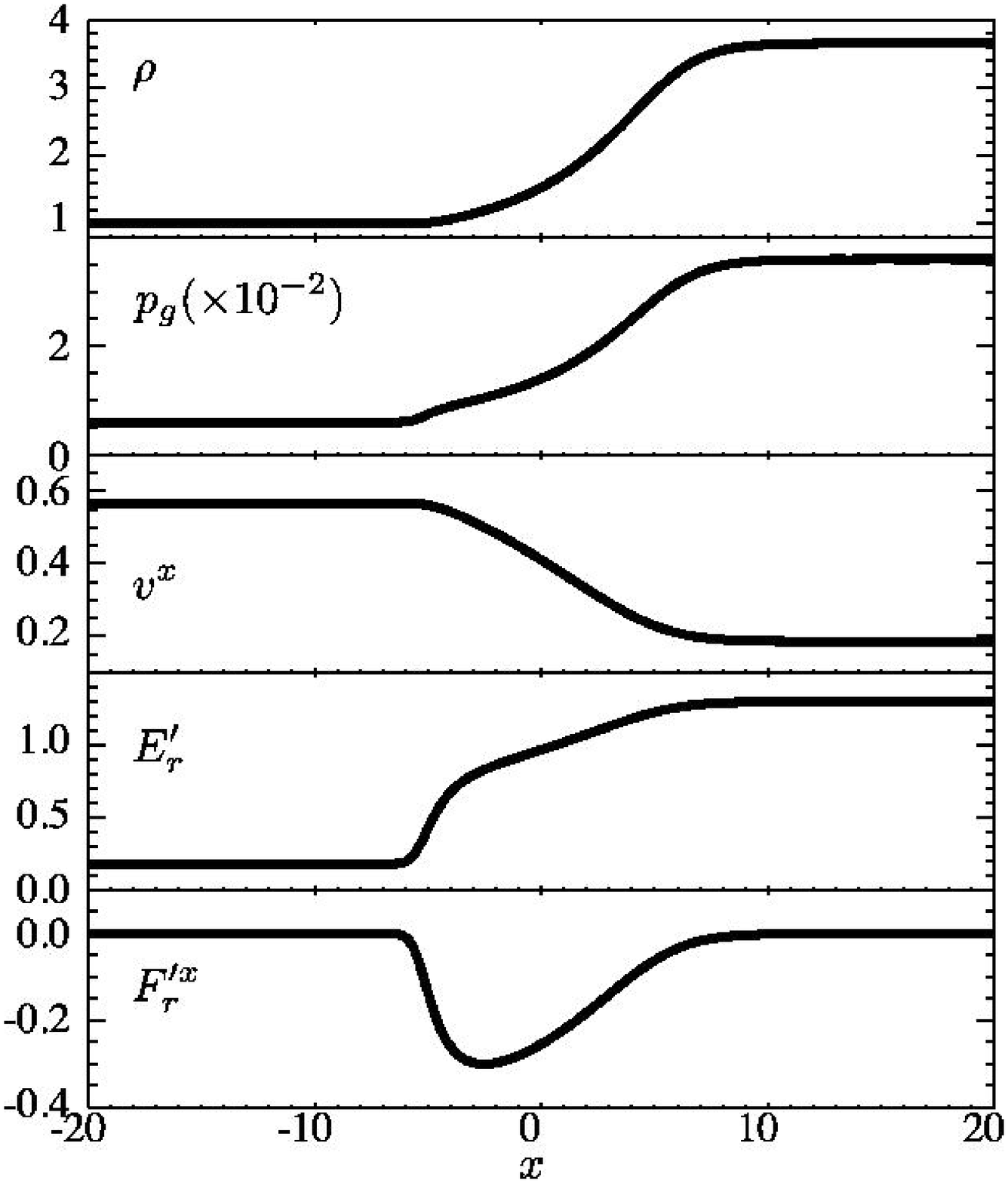}
   \caption{Profiles of $\rho$, $p_g$, $v^x$, $E'_r$ and $F'^x_r$ from
  top to bottom at $t=5000$ for the model RHDST4. 
Dots and solid curves denote for numerical and semi-analytical solutions.}
   \label{fig:farris4}
 \end{minipage}
  \hspace{1mm}
  \begin{minipage}{0.49\hsize}
   \includegraphics[height=9cm]{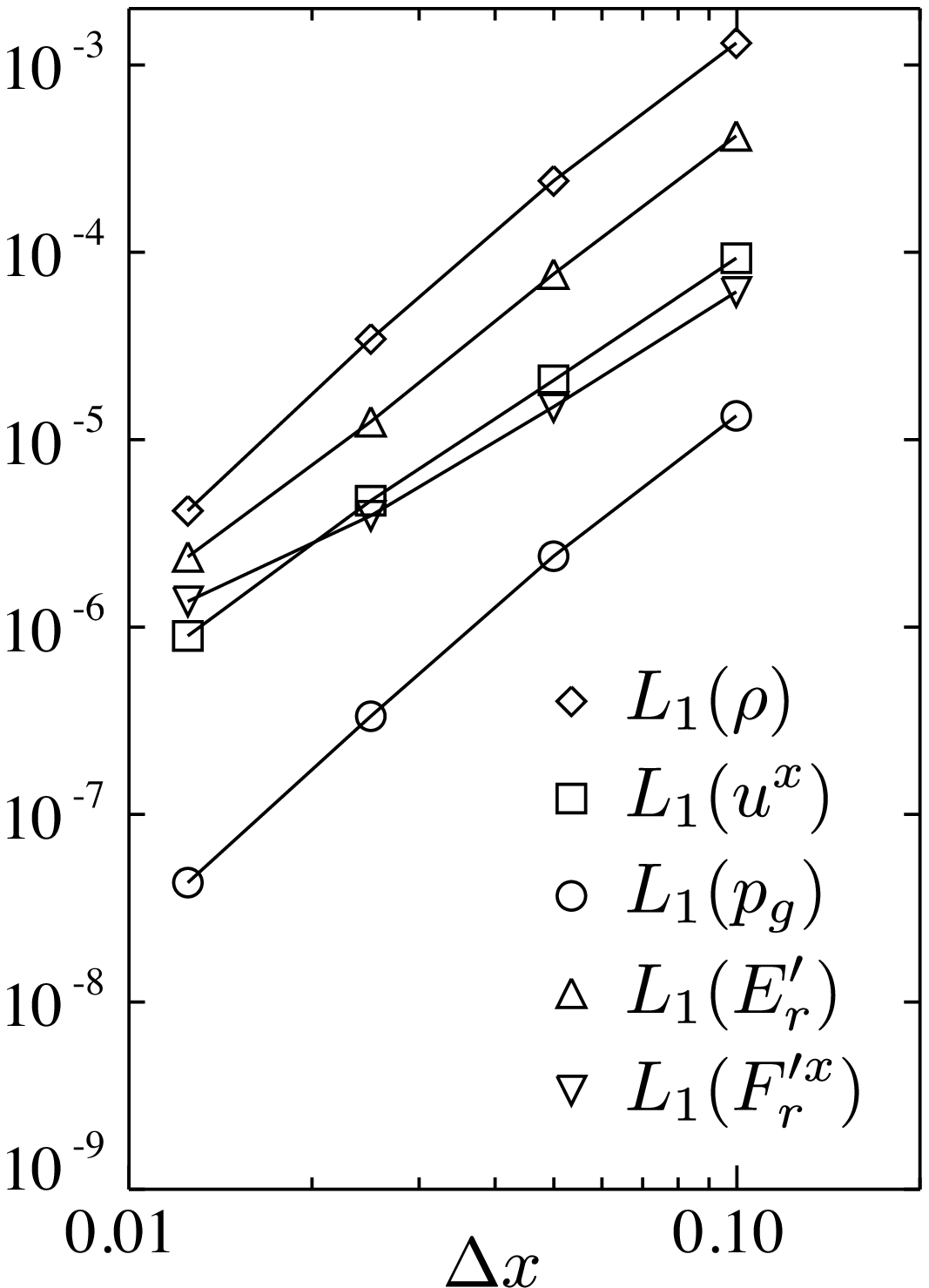}
   \caption{$L$-1 norms of $\rho$, $p_g$, $u^x$, $E'_r$ and
   $F'^x_r$ for the model RHDST4. }
   \label{fig:farris4-L1}
  \end{minipage}
 \end{tabular}
\end{figure*}

\subsubsection{Non-relativistic shock}\label{RHDST1}
In this test, the non-relativistic strong shock exists at $x=0$
(model RHDST1). The ratio of the radiation to thermal energy in the upstream
($x<0$) is $2.2\times 10^{-4}$. We take the CFL
condition to be $0.9$. 
Fig.~\ref{fig:farris1} shows profile at $t=5000$. The physical
quantities shown in this figure are $\rho$, $p_g$, $v^x$, $E_r', F'^x_r$
from top to bottom. (We note that $F'^{x}_r$, which is the
radiation flux measured in the comoving frame, is different from $F^x$
defined in \citealt{2008PhRvD..78b4023F} by factor $\gamma$).
Dots and curves denote for the numerical and semi-analytical solutions.

Since we give an initial condition by a step function at $x=0$, waves
arisen at the discontinuity propagate in the $\pm
x-$direction (mainly in the $+x$-direction since the upstream is
supersonic). After the waves pass the boundary at $x=\pm L$, the
system reaches the steady state. 

Since the radiation energy is negligible, fluid quantities
have jumps around $x=0$ similar to pure hydrodynamical shock. The
radiation field, on the other hand, has a
smooth profile in which the radiation energy is transported in front of the
shock. The radiation energy and flux have no discontinuities. 
The simulation results are well consistent with analytical solutions.

\subsubsection{Mildly relativistic shock}\label{RHDST2}
In this test, the mildly-relativistic strong shock exists at $x=0$
(model RHDST2). The ratio of the radiation to thermal energy in the upstream
is $3.3\times 10^{-3}$. We take the CFL
condition to be $0.9$. 
Fig.~\ref{fig:farris2} shows profiles at $t=5000$. The physical
quantities shown in this figure are $\rho$, $p_g$, $v^x$, $E_r', F'^x_r$ from top to bottom. 
We can see that the radiation energy density is no more continuous but
jumps at $x=0$. In the non-relativistic limit, the radiation energy
density and its flux are continuous, but it is not the case in
general. They are no more the conserved variables at the shock in the
relativistic flow \citep{2008PhRvD..78b4023F}. 
The solutions obtained in our numerical simulation (dots) 
are qualitatively and quantitatively consistent with semi-analytical
solutions (solid curves).

\subsubsection{Relativistic shock}\label{RHDST3}
In this test, the highly-relativistic strong shock exists at $x \simeq 0$
(model RHDST3). The upstream Lorentz factor is $\sim 10$ and the ratio
of the radiation to thermal energy is $3.3\times 10^{-2}$. We take the CFL
condition to be $0.9$. 
Fig.~\ref{fig:farris3} shows profile at $t=5000$. The physical
quantities shown in this figure are $\rho$, $p_g$, $v^x$, $E_r', F'^x_r$ from top to bottom. 

After waves generated at the jump $(x=0)$ passes the boundary, the
system approaches the steady state.
In this case, all the physical quantities are continuous.

The solutions obtained in our numerical simulation excellently recover
analytical solutions even when the flow speed is highly relativistic.   
To validate our numerical code, we compute the $L$-1 norms from
\begin{equation}
 L_1[f] = \Delta x \sum_{i=1}{N_x}|f_i - f_\mathrm{a}(x_i)|,
  \label{eq:L1}
\end{equation}
where $f_i$ is the physical quantity at the $i$-th grid and
$f_\mathrm{a}$ is the semi-analytic solution. 

We perform simulations with number of grid points $N_x=400, 800, 1600,
3200$. In these tests, we use the semi-analytic solution as the initial
condition. Figure~\ref{fig:farris3-L1} shows the $L$-1 norms of errors in
$\rho, u^x, p_g, E_r'$, and $F'^x_r$. We can see that all errors
converges at second order in $\Delta x$.

\subsubsection{Radiation dominated shock}\label{RHDST4}
In this test, the mildly-relativistic, radiation dominated shock exists
at $x\simeq 0$ (model RHDST4). The ratio
of the radiation to thermal energy in the upstream is $20$. We take
the CFL condition to be $0.3$. 
Fig.~\ref{fig:farris4} shows profile at $t=5000$. The physical
quantities shown in this figure are $\rho$, $p_g$, $v^x$, $E_r', F'^x_r$ from top to bottom. 

After waves generated at the jump $(x=0)$ pass the boundary, the
system approaches the steady state.
In this case, all the physical quantities are continuous and their
profiles are very smooth since a precursor generated from the shock
strongly affects the plasma.

We again find very good agreement with semi-analytical solutions even when
the radiation energy dominates the thermal energy. 
Figure~\ref{fig:farris4-L1} shows the $L$-1 norms of errors in $\rho, u^x,
p_g, E_r'$, and $F'^x_r$. In this test, we adopt the analytical solution
as an initial condition following \cite{2008PhRvD..78b4023F}.
We can see that all errors converges at second order in $\Delta x$.

\section{Discussion}\label{discussion}
In the current study, we construct the relativistic hydrodynamic
simulation code including the radiation field. The magnetic field, which
would play a
crucial role in the relativistic phenomena, is neglected for simplicity. 
It would be quiet simple to include the magnetic field using a well-developed
numerical scheme. When we extend our radiation hydrodynamic code to the
radiation magnetohydrodynamic one, the energy momentum tensor
$T_\mathrm{HD}^{\nu,\mu}$ is replaced by that of the magnetofluids, 
\begin{equation}
  T^{\mu\nu}_\mathrm{MHD} = (\rho \xi+b^2)u^\mu u^\nu - b^\mu b^\nu
   +\left(p_g + \frac{b^2}{2}\right)\eta^{\mu \nu},\label{geq:TMHD}
\end{equation}
where $b^\nu = \left\{\bmath u \cdot \bmath B, 
\left[\bmath B + (\bmath u \cdot\bmath B)\bmath u\right]/\gamma\right\}$ 
is the covariant form of the magnetic fields. 
The magnetic field evolves according to the induction equation for the
ideal MHD,
\begin{equation}
 \partial_\nu (u^\nu b^\mu - u^\mu b^\nu)=0.\label{geq:induction}
\end{equation}
Then the energy momentum conservation equation and the induction
equation are explicitly integrated using the HLL or higher order
approximate Riemann solvers \citep{2005MNRAS.364..126M,
2006MNRAS.368.1040M, 2007JCoPh.223..643H,2009MNRAS.393.1141M}.
The source term describing the interaction between the matter and the
radiation, is integrated by applying our proposed scheme. 

Another simplification made in this paper is adopting the Eddington
approximation that the radiation field is assumed to be isotropic in the
comoving frame. The wave velocity of
radiation field has a propagation speed of $c/\sqrt{3}$. This would
become an issue when the flow speed is relativistic ($v\simeq c$). The
flow speed, in principle, exceeds the phase speed of light mode so that
the radiation energy might accumulate in front of the plasma flow. 
Also the problem appears when we consider the magnetofluids. 
The fast magnetosonic wave speed can exceeds the reduced light
speed $c/\sqrt{3}$.
Another problem is that the radiation does not propagate in a straight
line since we assume that the radiation field is isotropic. When the optically
thick medium with a finite volume is irradiated from one side, there
should be a
shadow on the other side \citep{2003ApJS..147..197H}. Such a shadow is
no longer formed when we
utilize the Eddington approximation \citep{2007A&A...464..429G}. 
To overcome these problems, we should admit an anisotropy of radiation flux in
a comoving frame \citep{1976UCRL...78378L.CA,1978JQSRT..20..541M,
1984JQSRT..31..149L}. 
In our formulation, we do not specify the closure relation until
\S~\ref{closure}. Thus we can employ M-1 closure by replacing matrix
$A(u)$ without the other modification.
The explicit-implicit scheme of the relativistic radiation
magnetohydrodynamics with the M-1 closure will be reported in the near
future.

\section{Summary}\label{summary}
We have developed a numerical scheme for solving special relativistic
radiation hydrodynamics, which ensures the conservation of total
energy and flux. The hyperbolic term is explicitly solved using an
approximate Riemann solver, while the source term describing the
interaction between the matter and the radiation, is implicitly
integrated using an iteration method. 
The advantage of the implicit scheme is that 
we can take the numerical time step larger than the absorption and
scattering time scale.
This allows us to study the long term evolution of the system
(typically, the dynamical timescale).

When integrating the source term, we need to invert matrices $\bmath C$
and $\bmath A$ in our proposed scheme. 
We have to note that the rank of these matrices used in the implicit
scheme is very small ($4\times 4$ for $\bmath C$ and $6 \times 6$
for $\bmath A$). This is because the interaction between
the radiation and the gas is described by a local nature (source term)
when we solve 0th and 1st moments for radiation. 
This means that the numerical code can be easily parallelized and a high
parallelization efficiency is expected. 

We note that since the matrix $\bmath C$ is only $4\times 4$ matrix, we can invert it
analytically. For the matrix $\bmath A$, it is relatively small matrix but it
is difficult to invert it analytically. Thus we decide to use LU
decomposition method. Since the LU decomposition method can invert the 
matrix directly without iterations, we obtain $\bmath A^{-1}$ stably. 

We find that the wave front of radiation field can be sharply
captured using the HLL scheme even when we adopt the simple closure relation
$P'^{ij}_r=\delta^{ij}/3$ (i.e., the Eddington approximation).
When we adopt the FLD approximations, such a sharp wave front cannot be
captured due to the diffusion approximations. Thus, solving the 1st order
moment of radiation has an advantage when we consider the optically thin
medium although the wave speed reduces to $c/\sqrt{3}$ and the
radiation field is isotropic. 

We adopted the iteration scheme to integrate stiff source
terms. If we need many counts in iteration scheme, it causes a load
imbalance between each core when the code is parallelized. 
But we have to stress that solutions converged only within two steps
even if $\Delta t = 10^4 t_{ab}$ for the problem of radiative heating
and cooling appeared in \S~\ref{radheatcool}. For the other tests,
solution converges without iteration. Thus we can expect that the load
imbalance due to the iteration scheme is not severe even in the more 
realistic problems.

In the previous papers \citep{2008PhRvD..78b4023F, 2011MNRAS.tmp.1386Z},
radiation fields are defined in different frames for primitive and 
conservative variables (mixed frame). 
In such a method, the radiation moment equations are simply
described. However, this method is not suitable for implicit integration. 
This is because that a Lorentz transformation between two frames is
needed. By the transformation, expressions of the radiation
fields become quite complex even if the Eddington approximation is adopted. 
Moreover, the radiation stress tensor, Pr, is generally non-liner function 
of the radiation energy density and the radiation flux through the 
Eddington tensor. Thus, extension of explicit method to implicit method 
is not straightforward for the relativistic radiation hydrodynamics.

In our method, we treat radiation fields in the laboratory frame, only. 
Such a treatment simplifies implicit integration. Although we
need to invert $6\times 6$ matrix and $4\times 4$ matrix, our new method 
is easier than the method with mixed frame. Two matrices of $C$ 
(which is needed for implicit integration) and $A$ 
(which is needed to compute Pr) are directly inverted using analytical 
solution and LU-decomposition method. Since we need not to use some
iteration methods to invert them, our method is stable and comparatively
easy. Our method would be quite useful for numerical simulations of
relativistic astrophysical phenomena, e.g., black hole
accretion disks, relativistic jets, gamma-ray bursts, end so on, since
both the high density region and high velocity region exists, and
since the radiation processes are play important role for dynamics as
well as energetics.



\acknowledgments
We thank Tomoya Takiwaki for fruitful discussions.
Numerical computations were carried out on Cray XT4 at the Center for
Computational Astrophysics, CfCA, at the National Astronomical Observatory
of Japan, on Fujitsu FX-1 at the JAXA Supercomputer System (JSS) at the Japan
Aerospace Exploration Agency (JAXA), and on T2K at the University of
Tokyo. This work is supported in part by Ministry of Education, Culture,
Sports, Science, and Technology (MEXT) for Research Activity Start-up (HRT)
23840045, and for Young Scientist (KO) 20740115
(YS) 21018008, 21105511, 23740160 (TI) 22$\cdot$3369, 23740154.
KT is supported by the Research Fellowship from the Japan Society for
the Promotion of Science (JSPS) for Young Scientists.
A part of this research has been funded by MEXT HPCI STRATEGIC PROGRAM.


\appendix
\section{calculating $P_r^{ij}$}\label{apPr}
In this appendix, we show how to obtain the radiation stress in the
observer frame from $E_r$ and $F_r^i$ by assuming the Eddington
approximation. $P_r^{ij}$ is the $3\times3$ symmetric matrix and there
are six unknowns. Since equation (\ref{eq:Edclos}) is a linear with $E_r$
and $F_r^i$, we need to solve 6th order linear equations,
\begin{equation}
 \bmath A p = r,
\end{equation}
where $\bmath A=\{a_{ij}\}$, $p^T = (P_r^{11}, P_r^{22}, P_r^{33}, P_r^{12},
P_r^{13}, P_r^{23})$, and  $r^T = (R^{11}, R^{22}, R^{33}, R^{12},
R^{13}, R^{23})$. 
The explicit form of $a_{ij}$ is written as
\begin{eqnarray} a_{11}&=&
      1+(u_1)(u_1)\left[-\frac{1}{3}+\frac{(u^1)^2
		}{(1+\gamma)^2}\right]+\frac{2u^1u_1}{1+\gamma},\label{eq:a11}\\
a_{12}&=&
      (u_2)(u_2)\left[-\frac{1}{3}+\frac{(u^1)^2}{(1+\gamma)^2}\right],\label{eq:a12}\\
a_{13}&=&
      (u_3)(u_3)\left[-\frac{1}{3}+\frac{(u^1)^2}{(1+\gamma)^2}\right],\label{eq:a13}\\
a_{14}&=&
(u_1)(u_2)\left[-\frac{1}{3}+\frac{(u^1)^2}{(1+\gamma)^2}\right]+\frac{2u^1u_2}{1+\gamma},\label{eq:a14}\\
a_{15}&=&
(u_1)(u_3)\left[-\frac{1}{3}+\frac{(u^1)^2}{(1+\gamma)^2}\right]+\frac{2u^1u_3}{1+\gamma},\label{eq:a15}\\
a_{16}&=&
(u_2)(u_3)\left[-\frac{1}{3}+\frac{(u^1)^2}{(1+\gamma)^2}\right]\label{eq:a16}
\end{eqnarray}
\begin{eqnarray}a_{21}&=&
(u_1)(u_1)\left[-\frac{1}{3}+\frac{(u^2)^2}{(1+\gamma)^2}\right],\label{eq:a21}\\
a_{22}&=&
1+(u_2)(u_2)\left[-\frac{1}{3}+\frac{(u^2)^2}{(1+\gamma)^2}\right]\label{eq:a22}
+\frac{2u^2u_2}{1+\gamma},\\
a_{23}&=&
(u_3)(u_3)\left[-\frac{1}{3}+\frac{(u^2)^2}{(1+\gamma)^2}\right],\label{eq:a23}\\
a_{24}&=&
(u_1)(u_2)\left[-\frac{1}{3}+\frac{(u^2)^2}{(1+\gamma)^2}\right]
+\frac{2u^2u_1}{1+\gamma},\label{eq:a24}\\
a_{25}&=&
(u_1)(u_3)\left[-\frac{1}{3}+\frac{(u^2)^2}{(1+\gamma)^2}\right],\label{eq:a25}\\
a_{26}&=&
(u_2)(u_3)\left[-\frac{1}{3}+\frac{(u^2)^2}{(1+\gamma)^2}\right]
+\frac{2u^2u_3}{1+\gamma},\label{eq:a26}\\
\end{eqnarray}
\begin{eqnarray} a_{31}&=&
(u_1)(u_1)\left[-\frac{1}{3}+\frac{(u^3)^2}{(1+\gamma)^2}\right],\label{eq:a31}\\
a_{32}&=&
(u_2)(u_2)\left[-\frac{1}{3}+\frac{(u^3)^2}{(1+\gamma)^2}\right],\label{eq:a32}\\
a_{33}&=&
1+(u_3)(u_3)\left[-\frac{1}{3}+\frac{(u^3)^2}{(1+\gamma)^2}\right]
+\frac{2u^3u_3}{1+\gamma},\label{eq:a33}\\
a_{34}&=&
(u_1)(u_2)\left[-\frac{1}{3}+\frac{(u^3)^2}{(1+\gamma)^2}\right],\label{eq:a34}\\
a_{35}&=&
(u_1)(u_3)\left[-\frac{1}{3}+\frac{(u^3)^2}{(1+\gamma)^2}\right]
+\frac{2u^3u_1}{1+\gamma},\label{eq:a35}\\
a_{36}&=&
(u_2)(u_3)\left[-\frac{1}{3}+\frac{(u^3)^2}{(1+\gamma)^2}\right]
+\frac{2u^3u_2}{1+\gamma}\label{eq:a36},
\end{eqnarray}
\begin{eqnarray}   a_{41} &=& \frac{(u^1) (u^2) (u_1) (u_1)}{(1+\gamma)^2} 
 +\frac{(u^2) (u_1)}{1+\gamma},\label{eq:a41}\\
  a_{42} &=& \frac{(u^1) (u^2) (u_2) (u_2)}{(1+\gamma)^2} 
 +\frac{(u^1) (u_2)}{1+\gamma},\label{eq:a42}\\
  a_{43} &=& \frac{(u^1) (u^2) (u_3) (u_3)}{(1+\gamma)^2},\label{eq:a43}\\
  a_{44} &=& 1 + \frac{(u^1) (u^2) (u_1) (u_2)}{(1+\gamma)^2} 
 +\frac{(u^1) (u_1)+(u^2)(u_2)}{1+\gamma},\label{eq:a44}\\
  a_{45} &=& \frac{(u^1) (u^2) (u_1) (u_3)}{(1+\gamma)^2}
 +\frac{(u^2) (u_3)}{1+\gamma},\label{eq:a45}\\
  a_{46} &=& \frac{(u^1) (u^2) (u_2) (u_3)}{(1+\gamma)^2}
 +\frac{(u^1) (u_3)}{1+\gamma},\label{eq:a46}\\
\end{eqnarray}
\begin{eqnarray}   a_{51} &=& \frac{(u^1) (u^3) (u_1) (u_1)}{(1+\gamma)^2} 
 +\frac{(u^3) (u_1)}{1+\gamma},\label{eq:a51}\\
  a_{52} &=& \frac{(u^1) (u^3) (u_2) (u_2)}{(1+\gamma)^2},\label{eq:a52}\\ 
  a_{53} &=& \frac{(u^1) (u^3) (u_3) (u_3)}{(1+\gamma)^2}
 +\frac{(u^1) (u_3)}{1+\gamma},\label{eq:a53}\\
  a_{54} &=& \frac{(u^1) (u^3) (u_1) (u_2)}{(1+\gamma)^2} 
 +\frac{(u^3) (u_2)}{1+\gamma},\label{eq:a54}\\
  a_{55} &=& 1+\frac{(u^1) (u^3) (u_1) (u_3)}{(1+\gamma)^2}
 +\frac{(u^1) (u_1)+(u^3)(u_3)}{1+\gamma},\label{eq:a55}\\
  a_{56} &=& \frac{(u^1) (u^3) (u_2) (u_3)}{(1+\gamma)^2}
 +\frac{(u^1) (u_2)}{1+\gamma},\label{eq:a56}\\
\end{eqnarray}
\begin{eqnarray}   a_{61} &=& \frac{(u^2) (u^3) (u_1) (u_1)}{(1+\gamma)^2},\label{eq:a61}\\
  a_{62} &=& \frac{(u^2) (u^3) (u_2) (u_2)}{(1+\gamma)^2}
 +\frac{(u^3) (u_2)}{1+\gamma},\label{eq:a62}\\
  a_{63} &=& \frac{(u^2) (u^3) (u_3) (u_3)}{(1+\gamma)^2}
 +\frac{(u^2) (u_3)}{1+\gamma},\label{eq:a63}\\
  a_{64} &=& \frac{(u^2) (u^3) (u_1) (u_2)}{(1+\gamma)^2} 
 +\frac{(u^3) (u_1)}{1+\gamma},\label{eq:a64}\\
  a_{65} &=& \frac{(u^2) (u^3) (u_1) (u_3)}{(1+\gamma)^2}
 +\frac{(u^2)(u_1)}{1+\gamma},\label{eq:a65}\\
  a_{66} &=& 1+\frac{(u^2) (u^3) (u_2) (u_3)}{(1+\gamma)^2}
 +\frac{(u^2) (u_2)+(u^3)(u_3)}{1+\gamma}.\label{eq:a66}
\end{eqnarray}
In our numerical code, the matrix $\bmath A$ is inverted using
$LU$-decomposition method.

Also 
$(\partial P^{ij})/(\partial E_r)$ and
$(\partial P^{ij})/(\partial F_r^k)$ are obtained by solving the following equations
\begin{eqnarray}
 A \frac{\partial p}{\partial E_r} = \frac{\partial r}{\partial E_r},~
  A \frac{\partial p}{\partial F^k_r} = \frac{\partial r}{\partial F^k_r}, 
\end{eqnarray}
where, 
\begin{eqnarray}
 \frac{\partial R^{ij}}{\partial E_r}  
&=& \frac{\gamma^2}{3}\delta^{ij} - u^i u^j,\\
 \frac{\partial R^{ij}}{\partial F_r^k}  
  &=&-\frac{2}{3} \gamma\delta^{ij}u^k
  +u^i \delta^{j}_k + u^j\delta^{i}_k + \frac{2}{1+\gamma}u^i u^j u_k.
\end{eqnarray}
Since the matrix $A$ is a function of $u$ and independent of $E_r$ and
$F^k_r$, matrix $A^{-1}$ is computed once before calculating $P^{ij}_r,
(\partial P^{ij}_r)/(\partial E_r)$ and $(\partial
P_r^{ij})/(\partial F_r^k)$.


\end{document}